\documentclass{article}

\usepackage{blindtext} % Package to generate dummy text throughout this template 

\usepackage{mathptmx} % Use the Palatino font
\usepackage[T1]{fontenc} % Use 8-bit encoding that has 256 glyphs
\usepackage{microtype} % Slightly tweak font spacing for aesthetics

\usepackage[english]{babel} % Language hyphenation and typographical rules

\usepackage[hmarginratio=1:1,top=32mm,columnsep=20pt]{geometry} % Document margins
\usepackage[hang, small,labelfont=bf,up,textfont=it,up]{caption} % Custom captions under/above floats in tables or figures
\usepackage{booktabs} % Horizontal rules in tables

\usepackage{lettrine} % The lettrine is the first enlarged letter at the beginning of the text

\usepackage{enumitem} % Customized lists
\setlist[itemize]{noitemsep} % Make itemize lists more compact

\usepackage{abstract} % Allows abstract customization
\usepackage{titlesec} % Allows customization of titles
\titleformat{\section}[block]{\large\scshape}{\thesection.}{1em}{} % Change the look of the section titles
\titleformat{\subsection}[block]{\large}{\thesubsection.}{1em}{} % Change the look of the section titles

\usepackage{fancyhdr} % Headers and footers
\usepackage{titling} % Customizing the title section

\usepackage{hyperref} % For hyperlinks in the PDF
\usepackage{natbib}
\usepackage{graphicx}
\usepackage{array}

\pretitle{\begin{center}\Huge\bfseries} % Article title formatting
\posttitle{\end{center}} % Article title closing formatting
\title{Modelling Settling-Driven Gravitational Instabilities at the Base of Volcanic Clouds Using the Lattice Boltzmann Method} % Article title
\author{%
\textsc{Jonathan Lemus\textsuperscript{1,2*}, Allan Fries\textsuperscript{1}, Paul A. Jarvis\textsuperscript{1}, Costanza Bonadonna\textsuperscript{1},} \\[1ex] % Your name
\textsc{Bastien Chopard\textsuperscript{2}, Jonas Lätt\textsuperscript{2}} \\[1ex] % Your name
\normalsize \textsuperscript{1}Department of Earth Sciences, University of Geneva, 1205 Geneva, Switzerland \\ % Your institution
\normalsize \textsuperscript{2}Department of Computer Science, University of Geneva, 1227 Carouge, Switzerland \\ % Your institution
\normalsize \href{mailto:jonathan.lemus@unige.ch}{jonathan.lemus@unige.ch} % Your email address
}
\date{\today} % Leave empty to omit a date
\renewcommand{\maketitlehookd}{%
\begin{abstract}
\noindent Field observations and laboratory experiments have shown that ash sedimentation can be significantly affected by collective settling mechanisms that promote premature ash deposition, with important implications for associated impacts. Among these mechanisms, settling-driven gravitational instabilities result from the formation of a gravitationally-unstable particle boundary layer (PBL) that grows between volcanic ash clouds and the underlying atmosphere. The PBL destabilises once it reaches a critical thickness characterised by a dimensionless Grashof number, triggering the formation of rapid, downward-moving ash fingers that remain poorly characterised. We simulate this process by coupling a Lattice Boltzmann model, which solves the Navier-Stokes equations for the fluid phase, with a Weighted Essentially Non Oscillatory (WENO) finite difference scheme which solves the advection-diffusion-settling equation describing particle transport. Since the physical problem is advection dominated, the use of the WENO scheme reduces numerical diffusivity and ensures accurate tracking of the temporal evolution of the interface between the layers. We have validated the new model by showing that the simulated early-time growth rate of the instability is in very good agreement with that predicted by linear stability analysis, whilst the modelled late-stage behaviour also successfully reproduces quantitative results from published laboratory experiments. The results show that the model is capable of reproducing both the growth of the unstable PBL and the non-linear dependence of the fingers’ vertical velocity on both the initial particle concentration and the particle diameter. Our validated model is used to expand the parameter space explored experimentally and provides key insights into field studies. Our simulations reveal that the critical Grashof number for the instability is about ten times larger than expected by analogy with thermal convection. Moreover, as in the experiments, we found that instabilities do not develop above a given particle threshold. Finally, we quantify the evolution of the mass of particles deposited at the base of the numerical domain and demonstrate that the accumulation rate increases with time, while it is expected to be constant if particles settle individually. This suggests that real-time measurements of sedimentation rate from volcanic clouds may be able to distinguish finger sedimentation from individual particle settling.
\end{abstract}
}

\begin{document}

% Print the title
\maketitle

%----------------------------------------------------------------------------------------
%	ARTICLE CONTENTS
%----------------------------------------------------------------------------------------

\section{Introduction}

Explosive volcanic eruptions can inject large quantities of ash into the atmosphere, generating multiple hazards at various spatial and temporal scales \citep{Blong2000, Bonadonna2021a}. Subsequent volcanic ash dispersal and sedimentation can strongly disrupt air traffic \citep{Guffanti2008, Prata2015}, affect inhabited areas \citep{Jenkins2015, Spence2005}, and impact ecosystems and public health \citep{Gudmundsson2011, Wilson2011}. A good understanding of ash dispersal is critical for effective forecasting and management of the response to these hazards. Modern volcanic ash transport and dispersal models have now reached high levels of sophistication \citep{Bonadonna2012, Folch2012, Folch2020, Jones2007, Prata2021} but do not include all of the physical processes affecting ash transport, such as particle aggregation and settling-driven gravitational instabilities \citep[e.g.][]{Durant2015a}. Various studies have highlighted the need to take these processes into account by revealing discrepancies between field measurements and numerical models \citep{Scollo2008}, premature sedimentation of fine ash leading to bimodal grainsize distributions not only related to particle aggregation \citep{Bonadonna2011a, Manzella2015, Watt2015} and significant depletion of airborne fine ash close to the source \citep{Gouhier2019a}.

\begin{figure}[!h]
\centering
\includegraphics{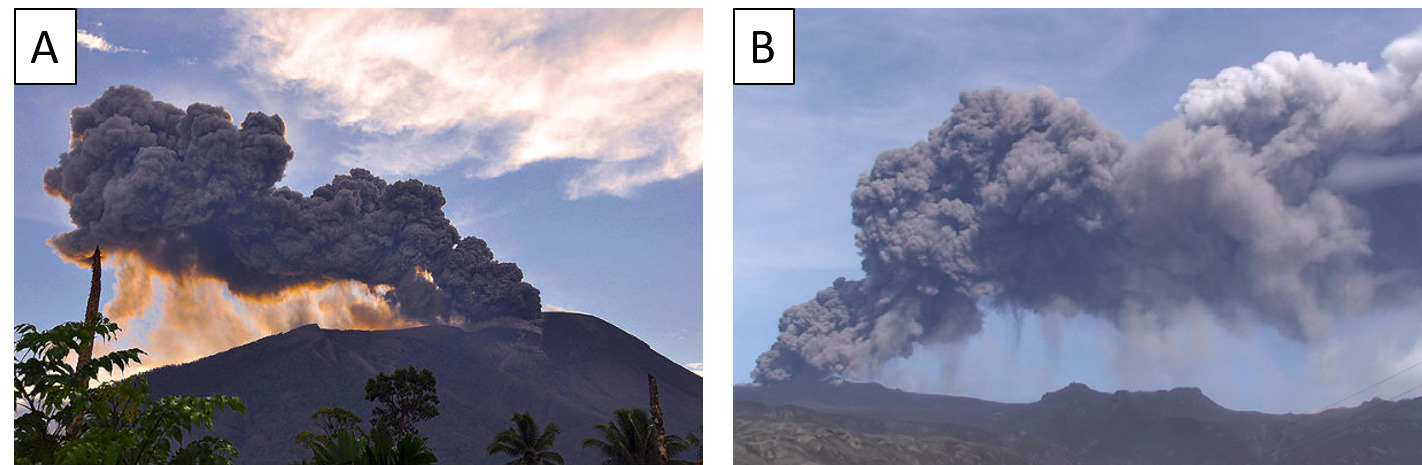}
\caption{Gravitational instabilities observed at the base of a volcanic plume during (A) the 2011 Gamalama eruption. (Credit: AP) and (B) the 2010 eruption of Eyjafjallajökull \citep{Manzella2015}}
\label{fig1}
\end{figure}

Alongside particle aggregation, settling-driven gravitational instabilities contribute to the early deposition of fine ash with similar outcomes (e.g. grainsize bimodality, premature sedimentation of fine ash). These instabilities generate downward-moving ash columns (fingers) which grow from the base of the ash cloud (Figure \ref{fig1}) \citep{Carazzo2012, Manzella2015, Scollo2017}. 

This phenomenon has the potential to enhance the sedimentation rate of fine ash beyond the terminal fall velocity of individual particles, reducing the residence time of fine ash in the atmosphere. Thus, a rigorous understanding of these processes is important in order to build a comprehensive parametrisation that can be included in dispersal models \citep{Bonadonna2012, Durant2015a, Folch2012, Scollo2010}.

Settling-driven gravitational instabilities occur at the interface between an upper, buoyant particle suspension, e.g., a volcanic ash cloud, and a lower, denser fluid, e.g., the underlying atmosphere \citep{Burns2012, DavarpanahJazi2016, Hoyal1999, Manzella2015}. Whilst the initial density configuration is stable, particle settling across the density interface creates a narrow unstable region called the particle boundary layer (PBL) \citep{Carazzo2012}. Once this attains a critical thickness \citep{Hoyal1999}, a Rayleigh-Taylor-like instability \citep{Chandrasekhar1961, Sharp1984} can form on the interface between the PBL and the lower layer, generating finger-like structures which propagate downwards. A further critical condition for instability is that the particle settling velocity $ V_s $ must be smaller than the finger propagation velocity $ V_f $ \citep{Carazzo2012}. Thus, the occurrence of the instability enhances the sedimentation rate  \citep{Manzella2015, Scollo2017}. Alternatively, if $ V_s $ is greater than the propagation velocity of fingers $ V_f $, then particles settle individually before a PBL can form and no instability occurs. Settling-driven gravitational instabilities have been widely studied in laboratory experiments that simulate various natural settings. Many experiments have considered an initial two-layer system, where the particle suspension is initially separated from the underlying denser layer by a removable horizontal barrier \citep{DavarpanahJazi2016, Fries2021, Harada2013, Hoyal1999, Manzella2015, Scollo2017} whilst other experiments have involved injection of the suspension into a density-stratified fluid at its neutral buoyancy level \citep{Carazzo2012, Cardoso2001a}. Similar instabilities can also be studied by allowing fine particles to sediment through the free surface between water and air \citep{Carey1997, Manville2004}. Additionally, dimensional analysis has been used to predict that the downward propagation velocity of the generated fingers is given by \citep{Carazzo2012, Hoyal1999}

\begin{equation}
V_f = \left[ g \left( \frac{\rho_{PBL} - \rho_f}{\rho_f} \right) \right]^{\frac{2}{5}} \left[ \frac{\pi V_s \delta^2_{PBL}}{4} \right]^{\frac{1}{5}},
\label{fing_vel}
\end{equation}
where $ \rho_{PBL} $ is the PBL bulk density, $ \rho_{f} $ the underlying fluid density, $g = 9.81 m.s^{-2}$ the gravitational acceleration and $ \delta_{PBL} $ the PBL thickness, which by analogy with thermal convection \citep{Turner1973} is taken to be \citep{Hoyal1999}

\begin{equation}
\delta_{PBL} = \left( \frac{Gr_c \nu^2}{g'} \right)^{\frac{1}{3}},
\label{delta_PBL}
\end{equation}
where $ g' = g \left( \rho_{PBL} - \rho_f \right) / \rho_f $, $ \nu $ the kinematic viscosity and $ Gr_c $ a critical Grashof number (see Table S1 in Supplementary Material for all acronyms and symbols used in this paper). The reduced gravity $ g' $ describes the change in the gravitational acceleration due to buoyancy forces.  Continuing the analogy with thermal convection, it has been proposed that $ Gr_c = 10^3 $ \citep{Hoyal1999}, although recent experimental observations suggests $ Gr_c \approx 10^4 $ may be more accurate \citep{Fries2021}. Therefore, for known particle and fluid properties, it is possible to predict whether collective settling will occur and fingers subsequently form using the condition $V_f \sup V_s$ \citep{Hoyal1999}. According to this relation, the limit between collective and individual settling occurs when $V_f = V_s$. However, the transition is likely to be smooth, with a transitionary regime where both fluid-like and particle-like settling occur at the same time, as suggested by \cite{Harada2013}.

For the initial two-layer configuration, \cite{Hoyal1999} also developed a series of analytical mass-balance models predicting the average particle concentration in the lower layer depending on whether the upper and lower layers were convecting or not. In the case of a quiescent upper layer and a convective lower layer (convection initiated by finger propagation), the evolution of the mass of particles in the lower layer $ M_2 $ depends on the balance between the mass flux of particles arriving from the upper layer $ \dot{M}_{in} $ and the mass flux of particle leaving by sedimentation $ \dot{M}_{out} $

\begin{equation}
\frac{dM_2}{dt} = \dot{M}_{in} - \dot{M}_{out},
\label{mass_bal}
\end{equation}
where $t$ is time. Assuming that $ M_2\left(t \right) = Ah_2C_2\left(t \right)$, where $C_2\left(t \right)$ is the average particle concentration in the lower layer, \cite{Hoyal1999} solved this equation using $ \dot{M}_{in} = AV_s C_1\left(0 \right)$, $ \dot{M}_{out} = AV_s C_2\left(t \right)$ and the initial condition $ C_2\left(0 \right) = 0$. Thus

\begin{equation}
C_2\left(t \right) = C_1\left(0 \right) \left[1 - e^{-\frac{V_s}{h_2}t} \right],
\label{C2_Hoy}
\end{equation}
where $C_1\left(0 \right)$  is the initial particle concentration in the upper layer, $h_2$ the lower layer thickness and $A$ the horizontal cross section of the tank.

Further studies of  settling-driven gravitational instabilities have taken theoretical approaches, such as using linear stability analyses to predict the growth rate and characteristic wavelengths of the instability at very early stages \citep{Alsinan2017, Burns2012, Yu2013}. Moreover, various numerical models simulating settling-driven gravitational instability have also been developed \citep{Burns2014, Chou2016, Jacobs2013, Keck2021, Yamamoto2015}. Most numerical approaches to this problem have used continuum-phase models, where the coupling between particles and fluid is strong enough to describe them as a single-phase \citep{Burns2014, Chou2016, XiaoYuTian-JianHsu2014}. This Eulerian description is valid under the assumptions of sufficiently small particles and a large enough number of particles such that the drag and gravitational forces are in equilibrium. The condition on the particle size can be quantified through the Stokes number \citep{burgisser:hal-00022567, Roche2019}

\begin{equation}
St = \frac{\rho_p D^2_p U}{18 \mu L},
\label{stokes}
\end{equation}
where $\rho_p$ is the particle density, $D_p$ the particle diameter, $\mu$ the dynamic viscosity and $U$ and $L$ characteristic velocity and length scales of the flow. For $St \inf 1$, the particles and fluid can be considered coupled and, providing there are enough particles, the continuum approach is valid.

The Eulerian description can be extended to multiple phases in order to simulate their interaction (e.g., gas-liquid interaction) using adaptive mesh refinements to resolve the phase interfaces \citep{Jacobs2013}. However, for large particle diameters and small particle volume fractions, collective behaviour no longer occurs and the continuum-phase method cannot be applied. In this case, there is a need to explicitly model particle motion, taking the drag force into consideration \citep{Chou2016, Yamamoto2015}. This paper presents an innovative method to implement a continuum model by coupling the Lattice Boltzmann Method (LBM) with a low-diffusivity finite difference (FD) scheme. This model takes advantage of the LBM capabilities to simulate complex flows through uniform grids and thus, the ease of coupling with finite difference methods. This hybrid model has been validated by comparing the results with those from linear stability analysis and laboratory experiments \citep{Fries2021}. The validated model then allows us to gain new insights into the fundamental processes by exploring experimentally-inaccessible regions of the parameter space. We first describe the general framework and governing equations that describe settling-driven gravitational instabilities, then the configuration of the validatory experiments to which we apply the model. Next, we propose a numerical strategy involving a hybrid model in order to solve the system of equations. We then go on to present the linear stability analysis before finally describing and discussing the results of our simulations.

%------------------------------------------------

\section{Methods}

\subsection{Problem formulation}
\label{prob_for}

The model consists of a three-way coupling between fluid momentum, fluid density, and particle volume fraction, based on the assumption that the particle suspension can be represented by a continuum concentration field. Moreover, the particle drag force is in equilibrium with the gravitational force such that the forcing term in the fluid momentum equation is equivalent to a buoyant force term (Boussinesq approximation), which depends on the particle volume fraction $\phi \left( \vec{x}, t \right)$ \citep{Burns2014, Chou2016, XiaoYuTian-JianHsu2014}. $\phi \left( \vec{x}, t \right)$ satisfies the advection-diffusion-settling equation

\begin{equation}
\frac{\partial \phi}{\partial t}  + \vec{\nabla}. [\phi (\vec{u}_f - V_s \vec{e}_z)] = D_c \nabla^2 \phi,
\label{ADS}
\end{equation}
where $\vec{u}_f \left(\vec{x}, t \right)$ is the fluid velocity, $D_c$ the particle diffusion coefficient, $\vec{e_z}$ the vertical unit vector and $\vec{x} = \left( x, y, z \right) $ the position coordinate. The fluid is considered incompressible meaning $ \vec{\nabla}. \vec{u}_f = 0 $. Thus, equation \ref{ADS} becomes

\begin{equation}
\frac{\partial \phi}{\partial t}  + (\vec{u}_f - V_s \vec{e}_z).\vec{\nabla} \phi - \phi \vec{\nabla}.(V_s \vec{e}_z) = D_c \nabla^2 \phi.
\label{ADS_inc}
\end{equation}

The particle settling velocity depends on the ambient fluid density $\rho$, which in turn depends on any transported density-altering properties, such as temperature or the concentration of a chemical species, e.g., the sugar in our validatory experiments \citep{Fries2021}. We incorporate the effect of a single density-altering property on the fluid density through a classical advection-diffusion equation 

\begin{equation}
\frac{\partial \rho \left( \rho_0, S \right)}{\partial t}  + \vec{u}_f.\vec{\nabla} \rho  \left( \rho_0, S \right) = D_s \nabla^2 \rho  \left( \rho_0, S \right),
\label{AD}
\end{equation}
where $\rho_0$ is a reference density of the carrier fluid, $S$ the density-altering quantity (temperature or concentration), and $D$ the associated diffusion coefficient. Additionally, under the Boussinesq approximation, we assume that the density depends linearly on $S$. The fluid momentum is modelled with the incompressible Navier-Stokes momentum equation

\begin{equation}
\frac{\partial \vec{u}_f}{\partial t}  + (\vec{u}_f.\vec{\nabla}) \vec{u}_f = - \frac{1}{\rho_0} \vec{\nabla} p + \nu \nabla^2 \vec{u}_f + \vec{F},
\label{mom}
\end{equation}
where $p$ is the pressure and $\vec{F}$ the buoyant body force term. We complete the system of equations by taking this force term to be a function of $\phi$ and $\rho$

\begin{equation}
\vec{F} = \left[\left(\frac{\rho_p - \rho_0}{\rho_0}\right) \phi + \left( \frac{\rho}{\rho_0} - 1 \right)(1 - \phi)\right]\vec{g}.
\label{forceterm}
\end{equation}

The system of equations presented so far assumes that all particles are of uniform size. In order to generalise to systems with polydisperse particle size distributions, we consider $N$ different particle concentration fields $\phi_i$, where each one is associated with a different size class and individually satisfies equation \ref{ADS_inc}. Furthermore, the body force term becomes

\begin{equation}
\vec{F} = \left[\left(\frac{\rho_p - \rho_0}{\rho_0}\right) \phi_{tot} + \left( \frac{\rho}{\rho_0} - 1 \right)(1 - \phi_{tot})\right]\vec{g}
\label{forceterm_poly}
\end{equation}
where

\begin{equation}
\phi_{tot} = \sum_{i=1}^{N} \phi_i .
\label{poly}
\end{equation}

\subsection{Flow configuration and experiment description}
\label{flow}

\begin{figure}[h]
\centering
\includegraphics[scale=0.8]{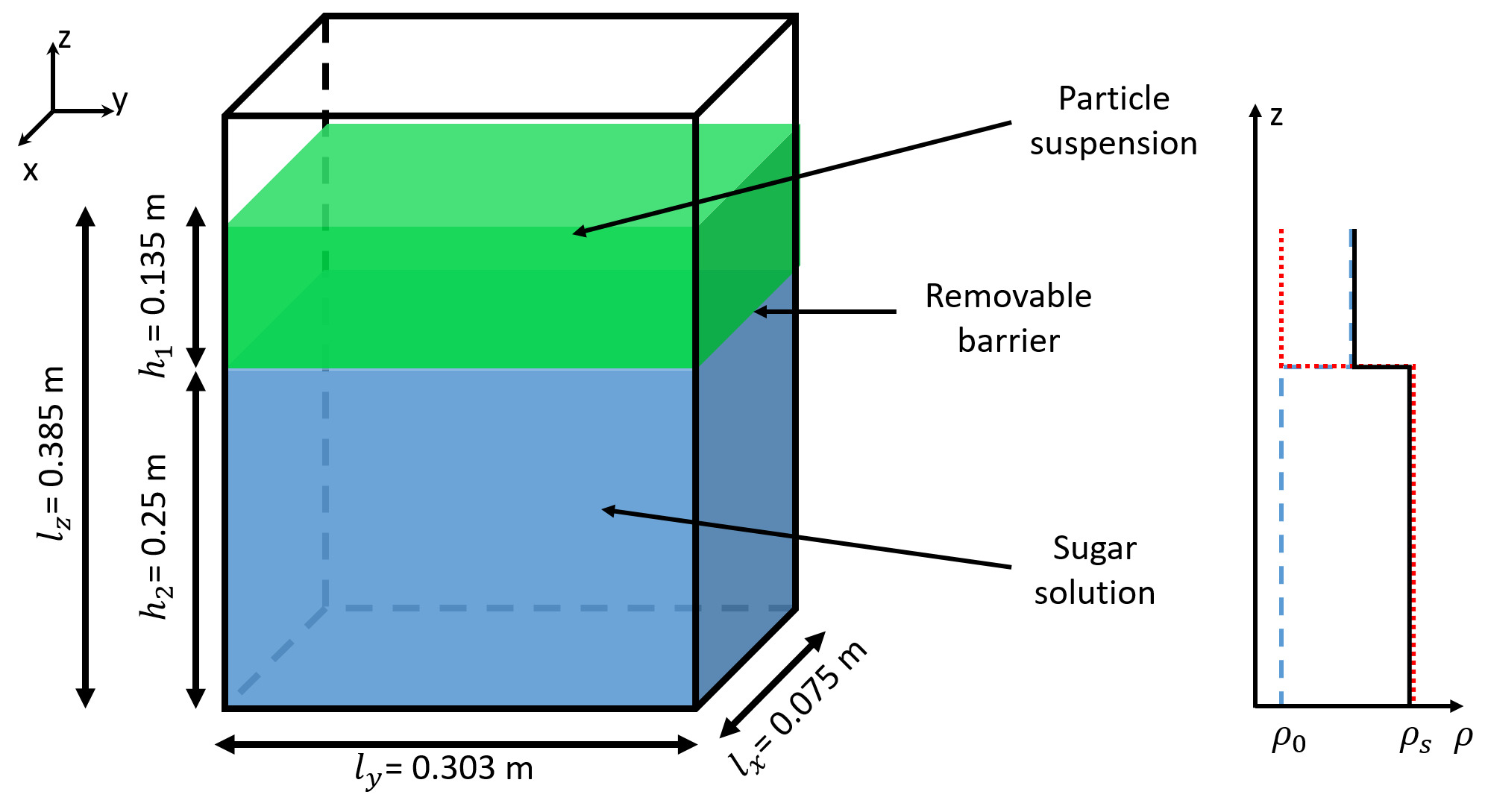}
\caption{Experimental setup used by \cite{Fries2021} and the initial density profiles associated with the contributions from particles (blue dashed) and sugar (red dotted), as well as the bulk density (black solid). The density of fresh water is given by $\rho_0$}
\label{fig2}
\end{figure}

Full details of the validatory laboratory experiments can be found in \cite{Fries2021} but we summarise the essential details here. The experiments are performed in a configuration identical to that of \cite{Manzella2015} and \cite{Scollo2017} (Figure \ref{fig2}) and consist of a water tank divided into two horizontal layers, initially separated by a removable barrier. The upper layer is an initially mixed particle suspension, which represents the ash cloud, and the lower layer is a dense sugar solution, analogue to the underlying atmosphere. The particles are spherical glass beads with a median diameter of $41.5 \pm 0.5 \, \mu m$ (measured using laser diffraction with a Bettersizer S3 Plus) and a density $\rho_p$ of $2519.4 \pm 0.09 \, kg/m^3$ (measured using helium pycnometry UltraPyc 1200e), and are sufficiently small to be well-coupled with the fluid, whilst the initial particle concentration $C_1 \left(0 \right)$ of the upper layer is varied from $1$ to $10 \, g/l$ (see Table \ref{table1} for the conversion to particle volume fraction $\phi_0$). The lower layer density is kept constant at $\rho_f = 1008.4 \, kg/m^3$ (corresponding to a sugar concentration of $S_0 = 35 \, g/l$), always ensuring an initially stable density configuration.

At the beginning of an experiment, the barrier separating the two layers is removed, allowing particle settling through the interface. A PBL subsequently forms and finger formation is initiated. Experiments are illuminated from the side of the water tank with a planar laser and recorded with a high-contrast camera. We measure the vertical finger velocity by tracking the progression of the finger front with time. Additionally, Planar Laser Induced Fluorescence (PLIF) \citep{Crimaldi2008, Koochesfahani1984} and particle imaging are used to quantify the spatial distribution of the fluid phase density and particle concentration.

\begin{table}[h!]
\centering
\begin{tabular}{|>{\centering\arraybackslash}m{2.5cm}|>{\centering\arraybackslash}m{1.8cm}|>{\centering\arraybackslash}m{1.5cm}|c|c|c|} 
 \hline
 Initial particle concentration $(g/l)$ & Volume fraction & Particle diameter $(\mu m)$ & $z_H \, (mm)$ & $z_{\phi} \, (mm)$ & $z_s \, (mm)$ \\ [0.5ex] 
 \hline\hline
 1 & $3.97 \times 10^{-4}$ & 40 & 12.11 & 2.59 & 0.67 \\
 2 & $7.94 \times 10^{-4}$ & 40 & 7.37 & 2.20 & 0.63 \\
 3 & $1.19 \times 10^{-3}$ & 40 & 5.47 & 1.99 & 0.61 \\ 
 4 & $1.59 \times 10^{-3}$ & 40 & 5.01 & 1.99 & 0.61 \\ 
 5 & $1.98 \times 10^{-3}$ & 40 & 4.20 & 1.84 & 0.60 \\ 
 6 & $2.38 \times 10^{-3}$ & 40 & 3.45 & 1.72 & 0.59 \\ 
 7 & $2.78 \times 10^{-3}$ & 40 & 3.25 & 1.70 & 0.60 \\ 
 8 & $3.18 \times 10^{-3}$ & 40 & 3.17 & 1.72 & 0.61\\ 
 9 & $3.57 \times 10^{-3}$ & 40 & 3.01 & 1.71 & 0.62 \\ 
 10 & $3.97 \times 10^{-3}$ & 40 & 2.54 & 1.58 & 0.61 \\ 
 3 & $1.19 \times 10^{-3}$ & 25 & - & - & - \\
 3 & $1.19 \times 10^{-3}$ & 55 & - & - & - \\
 3 & $1.19 \times 10^{-3}$ & 70 & - & - & - \\
 3 & $1.19 \times 10^{-3}$ & 85 & - & - & - \\
 3 & $1.19 \times 10^{-3}$ & 100 & - & - & - \\
 3 & $1.19 \times 10^{-3}$ & 115 & - & - & - \\
 3 & $1.19 \times 10^{-3}$ & 130 & - & - & - \\ 
 9 & $3.57 \times 10^{-3}$ & 25 & - & - & - \\ 
 9 & $3.57 \times 10^{-3}$ & 55 & - & - & - \\ 
 9 & $3.57 \times 10^{-3}$ & 70 & - & - & - \\ 
 9 & $3.57 \times 10^{-3}$ & 85 & - & - & - \\ 
 9 & $3.57 \times 10^{-3}$ & 100 & - & - & - \\ 
 9 & $3.57 \times 10^{-3}$ & 115 & - & - & - \\ 
 9 & $3.57 \times 10^{-3}$ & 130 & - & - & - \\ 
 9 & $3.57 \times 10^{-3}$ & 145 & - & - & - \\
 9 & $3.57 \times 10^{-3}$ & 160 & - & - & - \\
 [1ex] 
 \hline
\end{tabular}
\caption{List of simulations performed. All the simulations have been performed using an initial lower layer fluid density of $1008.4 \, kg/m^3$. $z_H$, $z_{\phi}$ and $z_s$ are parameters used in the linear stability analysis (LSA) in order to describe the different base states associated with the particle and sugar profiles in equations 39 and 40. The LSA has been performed only for a constant particle size of $40 \, \mu m$ in order to study the effect of the particle volume fraction.}
\label{table1}
\end{table}

\subsubsection{Application to flow configuration}
\label{appli}

We apply the general system of equations presented in section \ref{prob_for} to the configuration of the validatory experiments. The particles are spherical and sufficiently small that their terminal settling velocity in water is given by the Stokes velocity \citep{Stokes1851}

\begin{equation}
V_s = \frac{D^2_p g \left[ \rho_p - \rho \left(S \right) \right]}{18 \mu},
\label{stokes_vel}
\end{equation}
where $S$ is the sugar concentration and $\rho=  \rho_0 \left(1 + \alpha S \right)$, with $\alpha$ the sugar expansion coefficient.

We simulate the solid walls of the tank around our domain with a no-slip boundary condition for the fluid velocity. Neumann boundary conditions are employed for $\phi$ and $\rho$ to avoid any flux of particles or sugar across the walls. Thus we impose

\begin{equation}
\frac{\partial \phi}{\partial x} = 0,
\label{neum1}
\end{equation}
and

\begin{equation}
\frac{\partial \rho}{\partial x} = 0,
\label{neum2}
\end{equation}
on the wall nodes. Furthermore, we define the following initial states for $\phi$ and $S$

\begin{equation}
\phi \left( \vec{x}, t=0 \right) =     \left\{ \begin{array}{rcl}0  & \mbox{for} & z < H_0 \\\phi_0  & \mbox{for} & z > H_0\end{array}\right.
\label{ini1}
\end{equation}
and

\begin{equation}
S \left( \vec{x}, t=0 \right) =     \left\{ \begin{array}{rcl} S_0  & \mbox{for} & z < H_0 \\0  & \mbox{for} & z > H_0\end{array}\right.
\label{ini}
\end{equation}
where $\phi_0$ and $S_0$ are the initial particle volume fraction in the upper layer and initial sugar concentration in the lower layer, respectively, and $H_0=0.25 \, m$ the initial height of the interface ($z=0$ corresponds to the base of the tank). We also add a small perturbation to the particle volume fraction field in order to initiate the instability. Finally, the system is initially stationary so $\vec{u_f} \left( \vec{x} ,t=0 \right) =0$.

\subsection{Numerical methods}

The model is implemented using a hybrid strategy where a LBM solves the fluid motion and is coupled with finite difference schemes that solve the advection-diffusion equations for $\phi$ and $S$.

\subsubsection{Fluid motion}

The LBM is an efficient alternative to conventional Computational Fluid Dynamics (CFD) methods that explicitly solve the Navier-Stokes equations at each node of a discretised domain \citep{Chen2010, He1997}. It is a well-established approach for simulating complex flows, including multiphase fluids \citep{Leclaire2017} and thermal and buoyancy effects  \citep{Noriega_2013, Parmigiani2009}. The LBM originates from the kinetic theory of gases and provides a description of gas dynamics at the mesoscopic scale. This scale exists between the microscopic, which describes molecular dynamics, and the macroscopic, which gives a continuum description of the system with variables such as density and velocity. Thereby, the mesoscopic scale considers a probability distribution function of molecules described by the Lattice Boltzmann equation. This model reduces the process to two main steps: streaming (i.e., displacement of populations between consecutive calculation nodes), and collision (i.e., interaction of populations on a node). The Bhatnagar-Gross-Krook (BGK) model \citep{Bhatnagar1954} provides a simple collision process based on a fundamental property given by kinetic theory which describes gas motion as a perturbation around the equilibrium state. Then, the LBM-BGK model solves, for the particle population $f_i$, which are a discrete representation of the probability distribution function, the equation

\begin{equation}
f_i\left(\vec{x} + \vec{c_i} \delta t, t + \delta t \right) = -\frac{\delta t}{\tau} \left(f_i \left( \vec{x}, t \right) - f^{eq}_i \right),
\label{LBE}
\end{equation}
where $\delta t$ is the time step, $f^{eq}_i \left(\rho, \vec{u_f} \right)$ the equilibrium distribution function, $\tau$ the relaxation time associated with the flow viscosity and $\vec{c_i}$ the local particle velocity. The LBM is applied to specific types of lattices that describe how the populations move through the calculation nodes \citep{Kruger2017a}. These types of lattice are commonly summarized in the form $DrQm$ where $r$ denotes the dimension of the system and $m$ the number of directions in which populations can propagate.

\begin{figure}[h]
\centering
\includegraphics[scale=0.9]{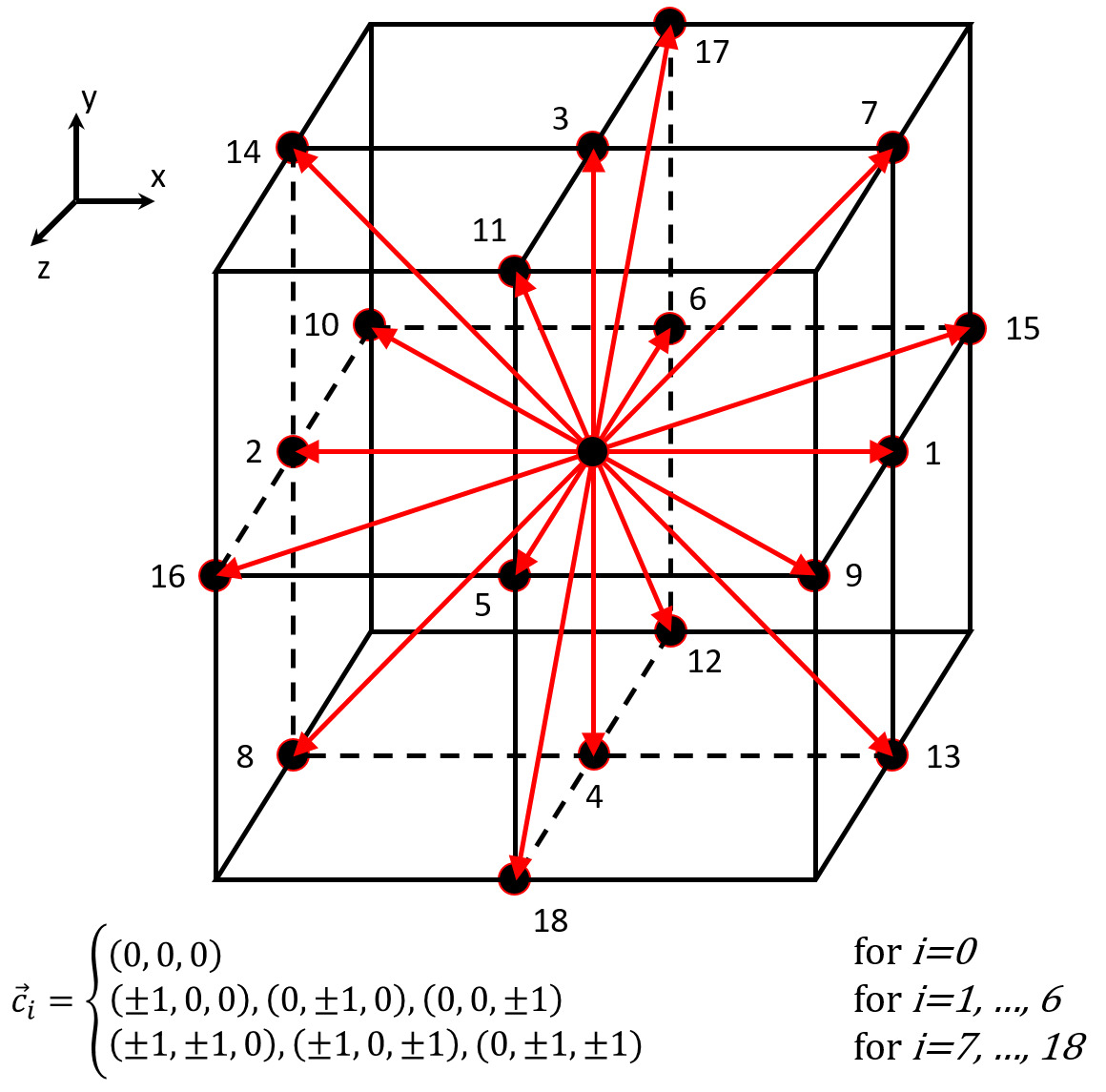}
\caption{Depiction of the D3Q19 lattice. The red arrows show the different possible directions of propagation. The associated local velocities are summarised in the velocity set $\vec{c_i}$.}
\label{fig3}
\end{figure}

Figure \ref{fig3} shows the scheme D3Q19 used for our 3D simulations and the associated set of local velocities. 

The macroscopic fluid state is described through the usual macroscopic variables such as density, velocity and kinematic viscosity. These variables are related to the moments of the populations $f_i$ through

\begin{equation}
\rho = \sum_i f_i,
\label{dens}
\end{equation}
and

\begin{equation}
\rho \vec{u_f} = \sum_i f_i \vec{c_i},
\label{mom_dens}
\end{equation}
whilst the kinematic viscosity controls the relaxation to equilibrium through the relaxation time

\begin{equation}
\tau = \frac{\nu}{c^2_s} + \frac{\delta t }{2},
\label{tau}
\end{equation}

The variable $c_s^2$  is commonly called the speed of sound and is equal to $(1/3) (\delta x^2 / \delta t^2)$  where $\delta x$ is the spatial step. However, the classical LBM-BGK model described above does not take into account any forcing term. One way to include forcing is to rewrite equation \ref{LBE} as

\begin{equation}
f_i\left(\vec{x} + \vec{c_i} \delta t, t + \delta t \right) = -\frac{\delta t}{\tau} \left(f_i \left( \vec{x}, t \right) - f^{eq}_i \right) + F_i \delta t,
\label{LBE_Guo}
\end{equation}
where $F_i$ is the forcing term, which can be expressed by a power series in the local particle velocities, and the equilibrium distribution is now given by $f^{eq}_i = f^{eq}_i \left(\rho, \vec{u^*_f} \right)$, where $\rho \vec{u^*_f} = \sum_i f_i \vec{c_i} + b\vec{F} \delta t$. The determination of the coefficient $b$, as well as the power series expansion of $F_i$ are described by \cite{Guo2002}. Finally, no-slip boundary conditions in the LBM, to simulate walls for example, can be implemented using the classical $bounce-back$ boundary condition \citep{Kruger2017a} where the populations arriving on a wall node during the streaming step are simply reflected back to their previous nodes.

\subsubsection{Transport of particles and other density-altering quantities}

The particles and other density-altering quantities are described by continuum fields that follow an advection-diffusion law coupled with the fluid motion as simulated with the LBM. The numerical solution of the advection equation is particularly challenging for methods which, like ours, are Eulerian (i.e., mesh-based). Indeed, such methods exhibit numerical diffusion which may strongly reduce model accuracy and, in some cases, even exceed the amplitude of the actual, physical diffusion term. The lack of physical diffusion in our problem and the presence of sharp interfaces restrict our ability to solve the advection equations with the LBM. In fact, the advection-diffusion equation can be solved by the LBM with a BGK approach in analogous fashion to the fluid motion by modifying the equilibrium distribution and the relaxation time to depend on the diffusion coefficient $D$ rather than $\nu$

\begin{equation}
\tau = \frac{D}{c^2_s} + \frac{\delta t }{2},
\label{tau2}
\end{equation}

However, a stability condition for a LBM-BGK algorithm is $\tau / \delta t = 1/2$. Thus, since the problem is convection dominated, the low diffusion coefficient ($D \ll 1$) drives the model towards the stability limit, introducing strong numerical errors near sharp concentration gradients \citep{Hosseini2017}. For this reason, we solve the advection term using two finite-difference schemes which are selected depending on the required accuracy: the classical first-order upwind finite difference and the third-order Weighted Essentially Non Oscillatory (WENO) finite difference scheme \citep{Jiang1996, Liu1994}.

Coupling the LBM with an upwind finite difference scheme allows us to avoid the stability problem. First-order FD schemes however, still suffer from the problem of numerical diffusion due to the truncation error associated with terminating the Taylor expansion after the first spatial derivative. The induced numerical error NE for the convective term in the advection-diffusion equation is given by

\begin{equation}
NE \sim u \frac{\delta x}{2} \frac{\partial^2 \phi }{\partial x^2},
\label{NE}
\end{equation}
where $u$ is the transport velocity. $NE$ acts like an additional diffusion term because of the presence of the second-order derivative. The numerical diffusion associated with the solution of $S$ is negligible due to the low fluid velocity and consequently the use of the first order FD scheme does not significantly affect the accuracy. However, in the solution of $\phi$, which includes an additional velocity contribution due to the settling, the truncation error associated with the first-order scheme becomes non-negligible. Whilst decreasing $\delta x$ would reduce numerical diffusion, we would require an unpractically small value in order to get a sufficiently accurate solution. Additionally, simply increasing the order of the scheme introduces dispersion (spurious oscillations) near regions of high gradient, according to the Godunov theorem \citep{godunov1954different, godunov1959difference}. Therefore, we choose here to implement the low diffusive WENO procedure for the solution of $\phi$, thus achieving a stable and high-resolution scheme without dispersion.

Further information on how we discretise the convective term in the advection-diffusion equation using the first order upwind and the third order WENO finite difference schemes is detailed in Section 1 of the Supplementary material.

\subsubsection{Numerical implementation}

Our model is implemented using $Palabos$ (Parallel Lattice Boltzmann Solver), a Computational Fluid Dynamics (CFD) solver based on the Lattice Boltzmann Method and developed by the Scientific Parallel Computing group of the Computer Science Department, University of Geneva \citep{Latt2020}. $Palabos$ is designed to perform calculations on massively parallel computers, thus allowing very small spatial resolutions in order to accurately simulate the finger dynamics.

\section{Linear stability analysis}

In order to validate our model, we compare the early-time simulated behaviour against predictions from linear stability analysis (LSA). LSA is applied to the onset of the physical instability at the interface between layers of different particle concentration. It involves defining a field equation-satisfying base state for each of the unknown fields in a problem and then applying an infinitesimally small perturbation to each of these fields. The equations are then expanded to linear order in the perturbation, with higher order terms assumed to be negligible. By assigning the perturbation to have the form of a complex waveform, the system of equations reduces to an eigenvalue problem, which can be solved to determine which wavelengths will grow or decay \citep{Chandrasekhar1961}. In this section, we assume that the system is invariant under translation in the x-y plane, thus reducing the analysis to a 2D problem. We strongly follow the procedure described by \citep{Burns2012} in order to solve our problem.

\subsection{Nondimensionalisation}
 
We nondimensionalise our system of equations by defining

\begin{equation}
l^c = \left( \frac{\nu^2}{g} \right)^\frac{1}{3},
\label{lc}
\end{equation}

\begin{equation}
t^c = \left( \frac{\nu}{g^2} \right)^\frac{1}{3},
\label{tc}
\end{equation}
and

\begin{equation}
p^c = \rho_0 \left( \nu g \right)^\frac{2}{3},
\label{pc}
\end{equation}
where $l^c$, $t^c$ and $p^c$ are characteristic quantities. We also define the dimensionless parameters

\begin{equation}
S^* = \alpha S_0,
\label{S_star}
\end{equation}

\begin{equation}
\phi^* = \phi_0,
\label{phi_star}
\end{equation}

\begin{equation}
Fr = \frac{1}{t^c} \sqrt{\frac{l^c}{g}},
\label{Fr}
\end{equation}
and

\begin{equation}
Sc_i = \frac{\nu}{D_i},
\label{Sc}
\end{equation}
noting that $Fr$ is a Froude number and $Sc_i$ are Schmidt numbers. Furthermore, the stream function $\psi$ is defined such that $\vec{u_f} = \left( \partial \psi / \partial z, -\partial \psi / \partial x \right)$  and the vorticity as $\vec{\omega} = \vec{\nabla} \times \vec{u_f}$. Then, applying the characteristic quantities to the vorticity formulation and equations (\ref{ADS_inc}-\ref{mom}), we obtain the dimensionless system (for the rest of the analysis, all the symbols used represent dimensionless quantities)

\begin{equation}
\omega = -\nabla^2 \psi,
\label{omega}
\end{equation}

\begin{equation}
\frac{\partial \omega}{\partial t} + \left(\vec{u_f}.\vec{\nabla}\right)\omega = \nabla^2 \omega + \frac{\partial \phi}{\partial y} \frac{\phi^*}{Fr^2} \left[ SS^* - \left( \frac{\rho_p - \rho_0}{\rho_0}\right)\right] - \frac{\partial S}{\partial y} \frac{S^*}{Fr^2} \left(1 - \phi \phi^*\right),
\label{omega2}
\end{equation}

\begin{equation}
\frac{\partial S}{\partial t} + \vec{u_f}.\vec{\nabla} S = \frac{1}{Sc_s} \nabla^2 S,
\label{S_nd}
\end{equation}
and

\begin{equation}
\frac{\partial \phi}{\partial t} + \left(\vec{u_f} - V_s \vec{e_z} \right).\vec{\nabla} \phi = \frac{1}{Sc_c} \nabla^2 \phi.
\label{phi_nd}
\end{equation}

Note that here we have neglected the term $-\phi \vec{\nabla}.\left(V_s \vec{e_z}\right)$ in equation \ref{ADS_inc} assuming the fluid density variation across the interface is sufficiently small that it does not affect the particle settling velocity.

\subsection{Variable expansion and eigenvalue problem}

We linearise the system of equations by expanding each variable in terms of a base state and a perturbation

\begin{equation}
\varphi \left(y, z, t \right) = \bar{\varphi} \left(z \right) + \varphi' \left(y, z, t \right),
\label{exp}
\end{equation}
where $\varphi \left(y, z, t \right) = \{ \psi, \omega, \phi, S \}$, $\bar{\varphi} \left(z \right) = \{ \bar{\psi}, \bar{\omega}, \bar{\phi}, \bar{S} \}$ the associated base state and $\varphi' \left(y, z, t \right) = \{ \psi', \omega', \phi', S' \}$ the perturbation. We choose the following base states

\begin{equation}
\bar{\psi} = 0,
\label{psi_bs}
\end{equation}

\begin{equation}
\bar{\omega} = 0,
\label{omega_bs}
\end{equation} 

\begin{equation}
\bar{\phi} \left( z, t \right) = \frac{1}{2} \left[ 1 + erf\left(\frac{z}{z_{\phi}} \right)\right],
\label{phi_bs}
\end{equation}

\begin{equation}
\bar{S} \left( z, t \right) = \frac{1}{2} \left[ 1 - erf\left(\frac{z-V_s t}{z_{S}} \right)\right],
\label{S_bs}
\end{equation}
where $ z_{\phi} \left( T \right) $ and $ z_{S} \left( T \right) $ are coefficients fitted in order to have similar base states to the profiles observed in the simulations prior to the onset of the instability which starts growing at the time $T$.

Solutions for the perturbation are assumed to have the form of normal modes 

\begin{equation}
\varphi' \left(y, z, t \right) = \hat{\varphi} \left( z \right)exp\left( iky + \sigma t \right),
\label{normodes}
\end{equation}
where $\hat{\varphi} \left( z \right)$ is the perturbation amplitude, $k$ the wavenumber and $\sigma$ the instability growth rate. The linearised system of equations is then formulated in matrix form so that the problem is reduced to the eigenvalue problem $K \vec{x} = \sigma W \vec{x}$ where

\begin{equation}
\vec{x} = \left( \begin{array}{c} \hat{\psi}\left( z \right)  \\ \hat{\omega}\left( z \right) \\ \hat{S}\left( z \right) \\ \hat{\phi}\left( z \right) \end{array} \right),
\label{vecmodes}
\end{equation}
and, in a reference frame moving downward at $V_s$, the matrices $K$ and $W$ are given by

\begin{equation}
K = \left( \begin{array}{c c c c} M & I & 0 & 0 \\
0 & M-V_s D_z & -ik\frac{S^*}{Fr^2} \left( 1 - \bar{\phi} \phi^* \right) I & ik \frac{\phi^*}{Fr^2} \left[ SS^* - \left( \frac{\rho_p - \rho_0}{\rho_0} \right) \right] I \\
ik \frac{d\bar{S}}{dz} I & 0 & \frac{1}{Sc_S} M - V_s D_z & 0 \\
ik \frac{d \bar{\phi}}{dz} I & 0 & 0 & \frac{1}{Sc_c} M \end{array} \right),
\label{matrix1}
\end{equation}
and

\begin{equation}
W = \left( \begin{array}{c c c c} 0 & 0 & 0 & 0 \\
0 & I & 0 & 0 \\
0 & 0 & I & 0 \\
0 & 0 & 0 & I \end{array} \right),
\label{matrix2}
\end{equation}
where $D_z = \partial / \partial z$, $M = -k^2 + D^2_z$ and $I$ is the identity operator.

The eigenvalues $\sigma$ determine the stability of the system:

\begin{itemize}
\item If all the eigenvalues have negative real parts, the system remains stable
\item If at least one eigenvalue has a positive real part, the system is unstable.
\end{itemize}

In order to solve the eigenvalue problem, the spatial derivatives are discretised using the linear rational collocation method with a grid transformation allowing a fine resolution around narrow interfacial regions \citep{Baltensperger2001, Berrut2004}.

\begin{figure}[h]
\centering
\includegraphics[scale=0.9]{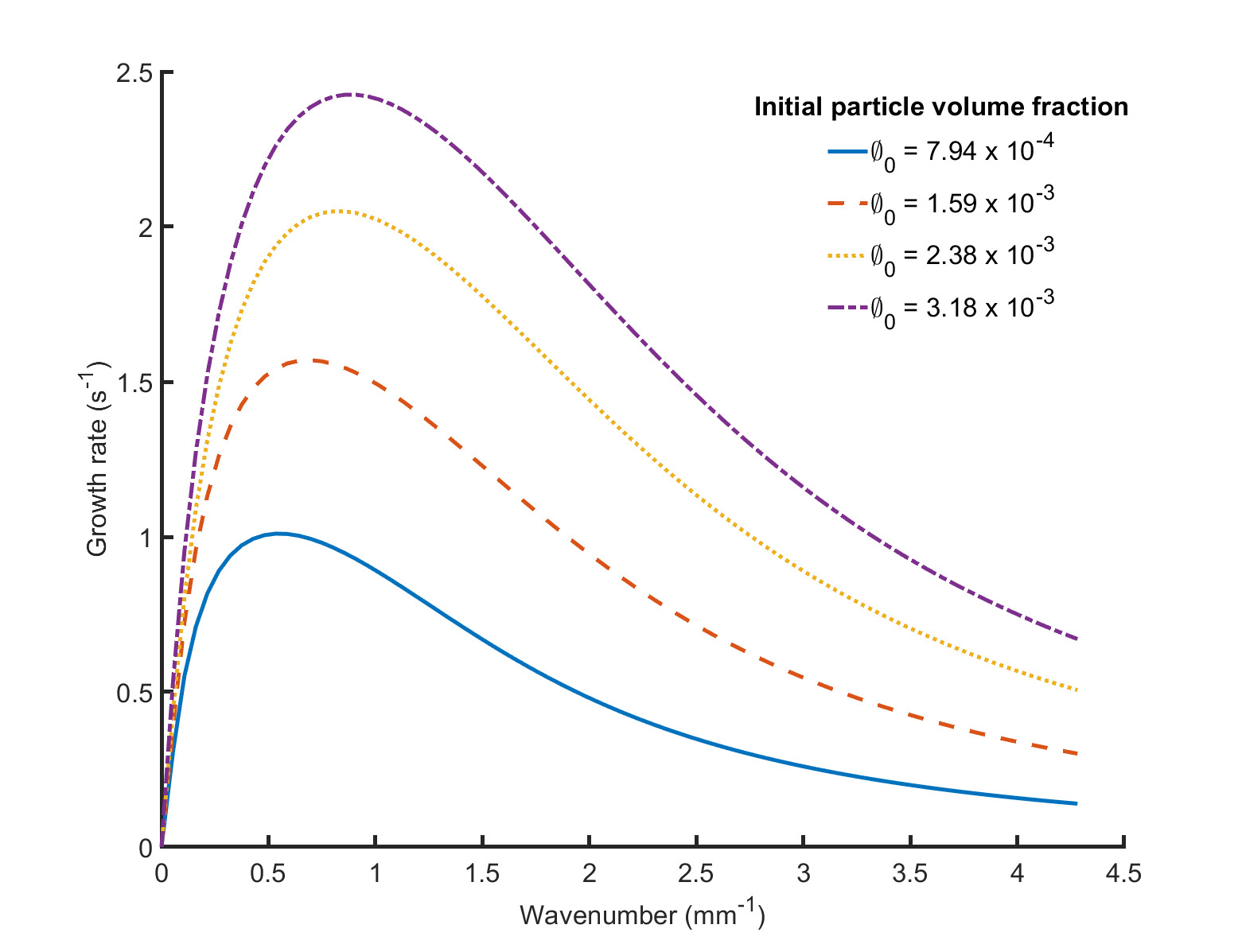}
\caption{Dispersion relation obtained from LSA for several initial particle concentrations.}
\label{fig4}
\end{figure}

The key result of the LSA is the dispersion relation between $\sigma$ and $k$. Figure \ref{fig4} presents the growth rate as a function of the wavenumber, for different initial particle volume fractions. The parameters for the different base states used to produce these curves are summarised in the Table \ref{table1}. We use this result in Section \ref{res_LSA} in order to compare the predictions of the LSA with the results of our numerical model.

%------------------------------------------------

\section{Results}

We validate our numerical model by comparing the results with predictions from LSA and experimental observations. The LSA predicts the growth rates of different perturbation wavenumbers during the very early stage of the instability, which can be compared with the spectrum of wavenumbers present in the particle concentration interface in the numerical model. Additionally, the experiments of \cite{Fries2021} employ imaging techniques to measure quantities, such as the particle concentration field and finger velocity, at times beyond the linear regime. Finally, our results are compared with some results of previous analyses on settling-driven gravitational instabilities \citep{Carazzo2012, Hoyal1999}.

\subsection{Comparison of model results with predictions from linear stability analysis}
\label{res_LSA}

\begin{figure}[h]
\centering
\includegraphics[scale=0.9]{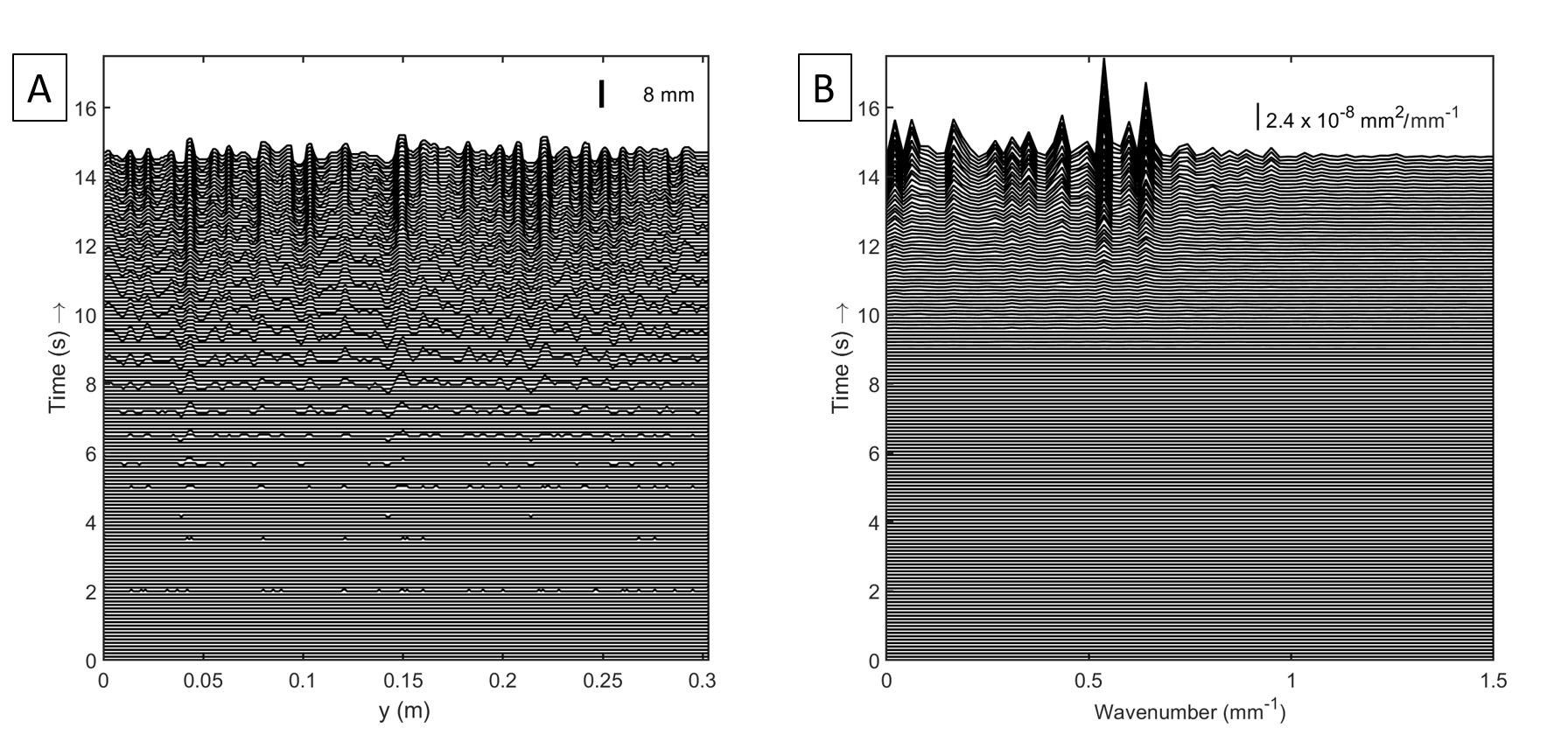}
\caption{(A) Space-time diagram of the particle front height $H\left(y, t \right)$. (B) Evolution of the power spectral density of the particle interface over time. Initial particle volume fraction: $\phi_0 =3.97 \times 10^{-4}$.}
\label{fig5}
\end{figure}

In order to compare our 3D simulations with the 2D linear stability analysis, we consider just the central plane of the simulation domain, i.e., a slice in the $\left( y, z \right)$ plane located at $x = l_x / 2$ ($l_x$ being the tank depth) (Figure \ref{fig2}). We define the front of the particle field to be the lowest position where $\phi= \phi_0 /2$ and also define $H\left(y \right)$ to be the separation between $z=0$ and this front. Our study has shown that the front position is only weakly affected when using other possible thresholds, i.e., $\phi_0 /10$ or $\phi_0 /5$ (relative change $~3\%$). Figure \ref{fig5}A shows an example of a space-time diagram showing the evolution of $H\left(y \right)$ through time. Furthermore, by calculating the Fourier transform $\tilde{H}\left(k, t \right)$ of $H\left(y \right)$ at different times, we can identify different dominant wavenumbers and their associated amplitudes as shown in the space diagram of the power spectral density (PSD) $\Gamma_H \left(k, t \right) = \left(1 / \left( k_S L_S \right) \right) |\tilde{H}\left(k, t \right)|^2 $ (Figure \ref{fig5}B), where $k_S$ is the sampling wavenumber and $L_S$ the number of samples. We extract the dominant mode and its associated growth rate from $ \Gamma_H \left(k, t \right) $ and compare the results with the predicted growth rates from LSA. We apply this analysis during a period when the amplitude $ |\tilde{H}\left(k, t \right)| $ of any given mode does not exceed $40 \%$ of its wavelength, thus ensuring we are still in the linear regime \citep{Lewis1950}.

During the linear regime, we can assume that the growth of the spectral amplitude can be described as \citep{Voltz2002}

\begin{equation}
 |\tilde{H}\left(k, t \right)| =   |\tilde{H}_i\left(k \right)| exp\left( \sigma_{sim}\left( k \right) t \right),
\label{Fourier}
\end{equation}
with $|\tilde{H}_i\left(k \right)|$ the initial amplitude and $\sigma_{sim}\left( k \right)$ the instability growth rate as determined from the simulations. Thus, the PSD can be expressed as

\begin{equation}
 \Gamma_H \left(k, t \right) =   \Gamma_{H_i} exp\left(2 \sigma_{sim}\left( k \right) t \right),
\label{PSD}
\end{equation}
where $\Gamma_{H_i}$ is the initial spectral density. 

\begin{figure}[h]
\centering
\includegraphics[scale=0.9]{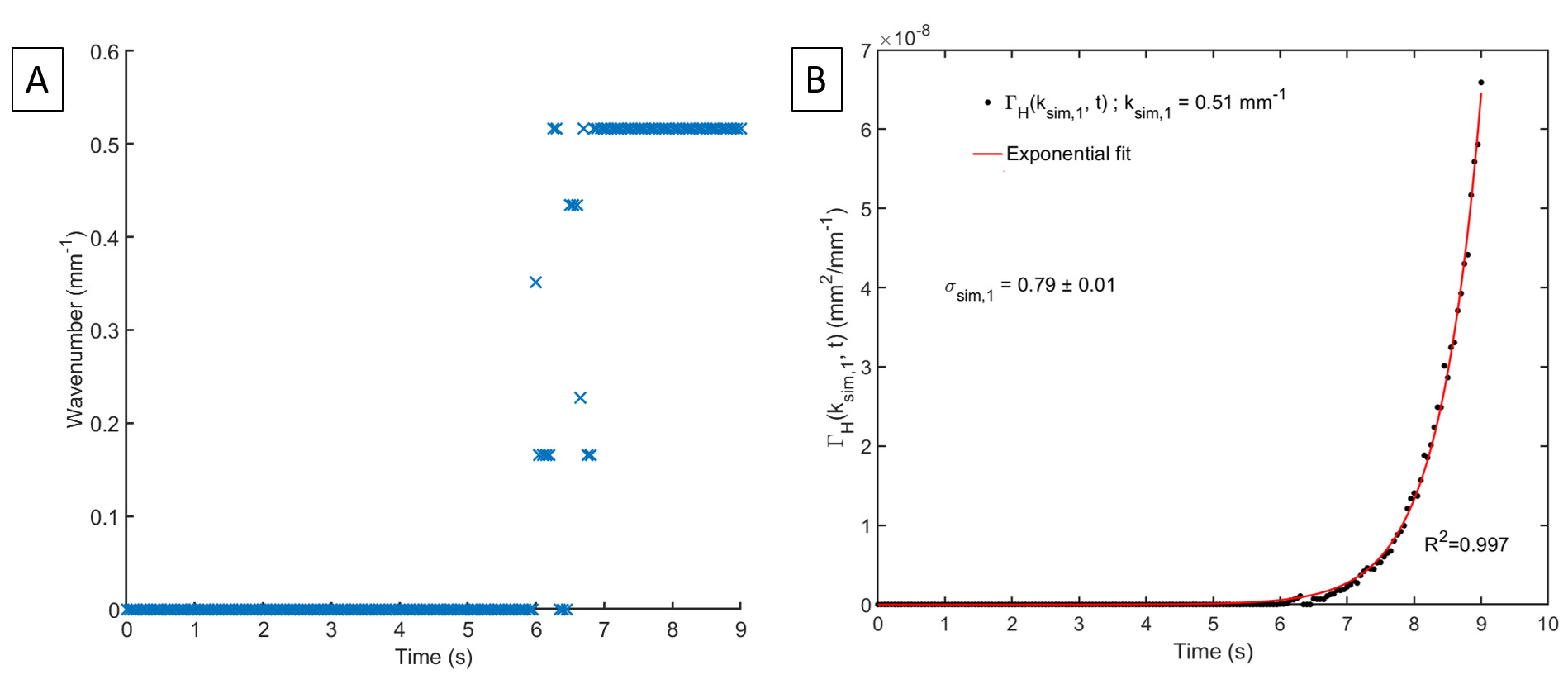}
\caption{(A) Example of dominant wavenumbers extracted from the maximum of the PSD. Initial particle volume fraction $\phi_0= 7.94 \times 10^{-4}$. (B) Exponential fitting to the temporal evolution of the PSD for the first maximum in (A), $k_{sim,1} = 0.517 \, mm^{-1}$.}
\label{fig6}
\end{figure}

At each time step, we extract the PSD and the wavenumber $k_{sim}$ associated with the dominant mode as shown in Figure \ref{fig6}. However, we observe that the dominant mode remains at the same wavenumber during instability growth except for three cases ($\phi_0 = 1.59 \times 10^{-3}$, $2.38 \times 10^{-3}$ and $3.97 \times 10^{-3}$) where we observed that the dominant mode changed its position in the spectral space. For these simulations only, we have a set of several wavenumbers $k_{sim,i}$, ($i=1,2,3$) associated with the dominant mode. With the computed PSD of the dominant mode as a function of time $\Gamma_H \left( k_{sim},t \right)$, we apply our exponential fitting (equation \ref{PSD}) to determine the growth rate $\sigma_{sim,i}$ (Figure \ref{fig6}B). For the simulations which resulted in several values of $k_{sim,i}$ for the dominant mode, we measured the growth rates of each mode $\sigma_{sim,i}$ and found identical values, up to a precision of $5\%$. Additionally, for each simulation, we find the time $T$ when the instability starts growing, easily identified as the time at which the modal wavenumber becomes non-zero (e.g., in Figure \ref{fig6}A this is at approximately 6 s). At this time, we extract the associated vertical profiles of particle and sugar concentration which are used to find the coefficients $z_{\phi} (T)$ and $z_S (T)$ (equations \ref{phi_bs}-\ref{S_bs}) and thus determine the base states of $\bar{\phi}(z,T)$ and $\bar{S}(z,T)$ (Figure 7). We then perform the LSA for each $\phi_0$, using the appropriate base states, and obtain a dispersion relation $\sigma=f(k)$. Using this relation, we predict the different growth rates  $\sigma=\sigma_{LSA,i}$ associated with $k=k_{sim,i}$ and we compare with $\sigma (k_{sim,i})$ as measured in our simulations. Figure 8 shows the comparison between $\sigma_{sim}$ (black dots) and $\sigma_{LSA,i}$ (red triangles), as predicted from the LSA, for the dominant wave mode. For the cases including a moving dominant mode, we plotted the growth rates associated with the different measured wavenumbers. We see that the dependence of the largest value of $\sigma_{LSA,i}$ on the initial particle concentration is in good agreement with the simulated growth rate.

\begin{figure}[h]
\centering
\includegraphics[scale=0.9]{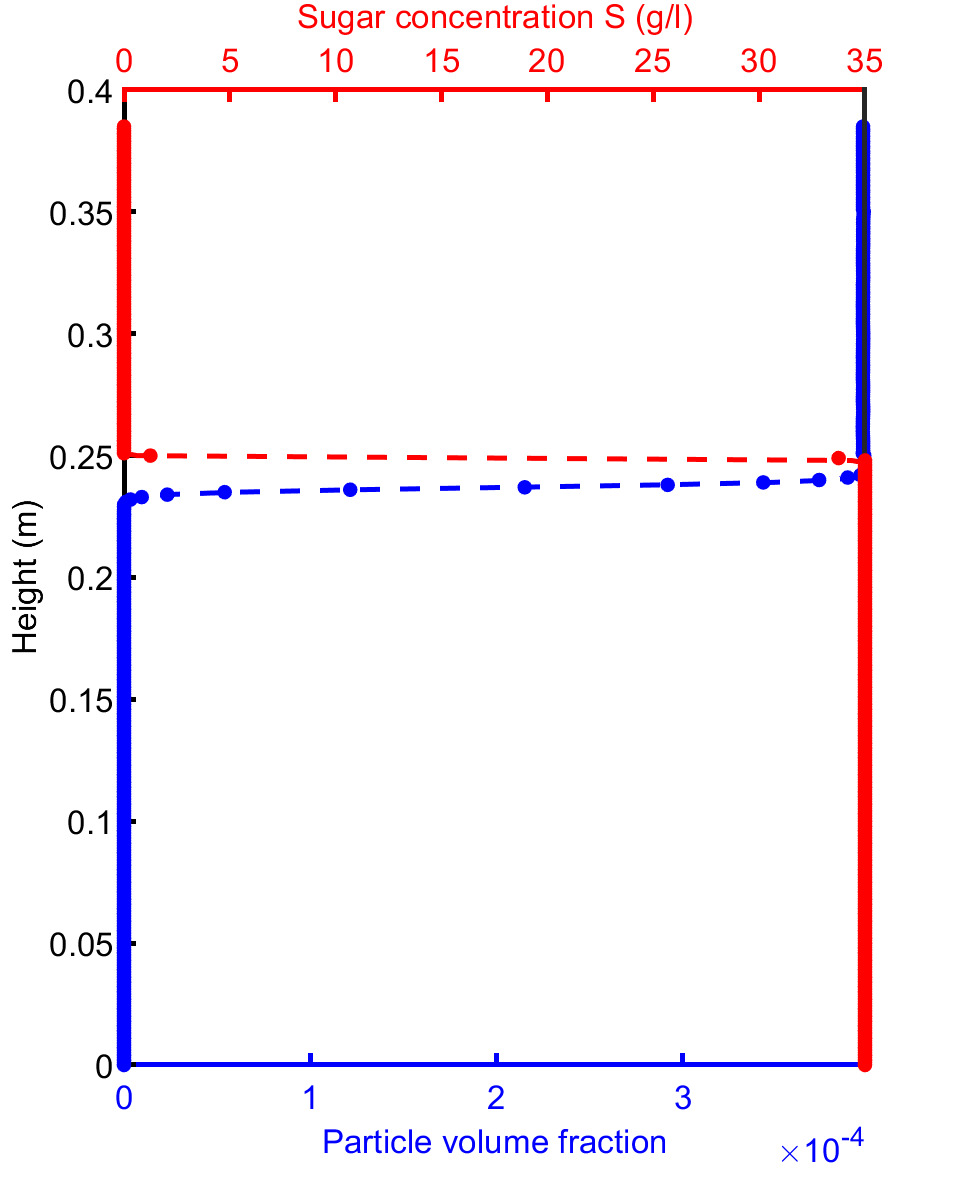}
\caption{Example of LSA base states extracted from the simulations for $\phi_0 = 3.97 \times 10^{-4}$. Dots: profiles extracted from the simulations at $T = 9.55s$ (start of the instability growth). Dotted lines: Fit with equations \ref{phi_bs} and \ref{S_bs}. i.e., base states used for the LSA. Blue: particle volume fraction. Red: Sugar concentration.}
\label{fig7}
\end{figure}

\begin{figure}[h]
\centering
\includegraphics[scale=0.9]{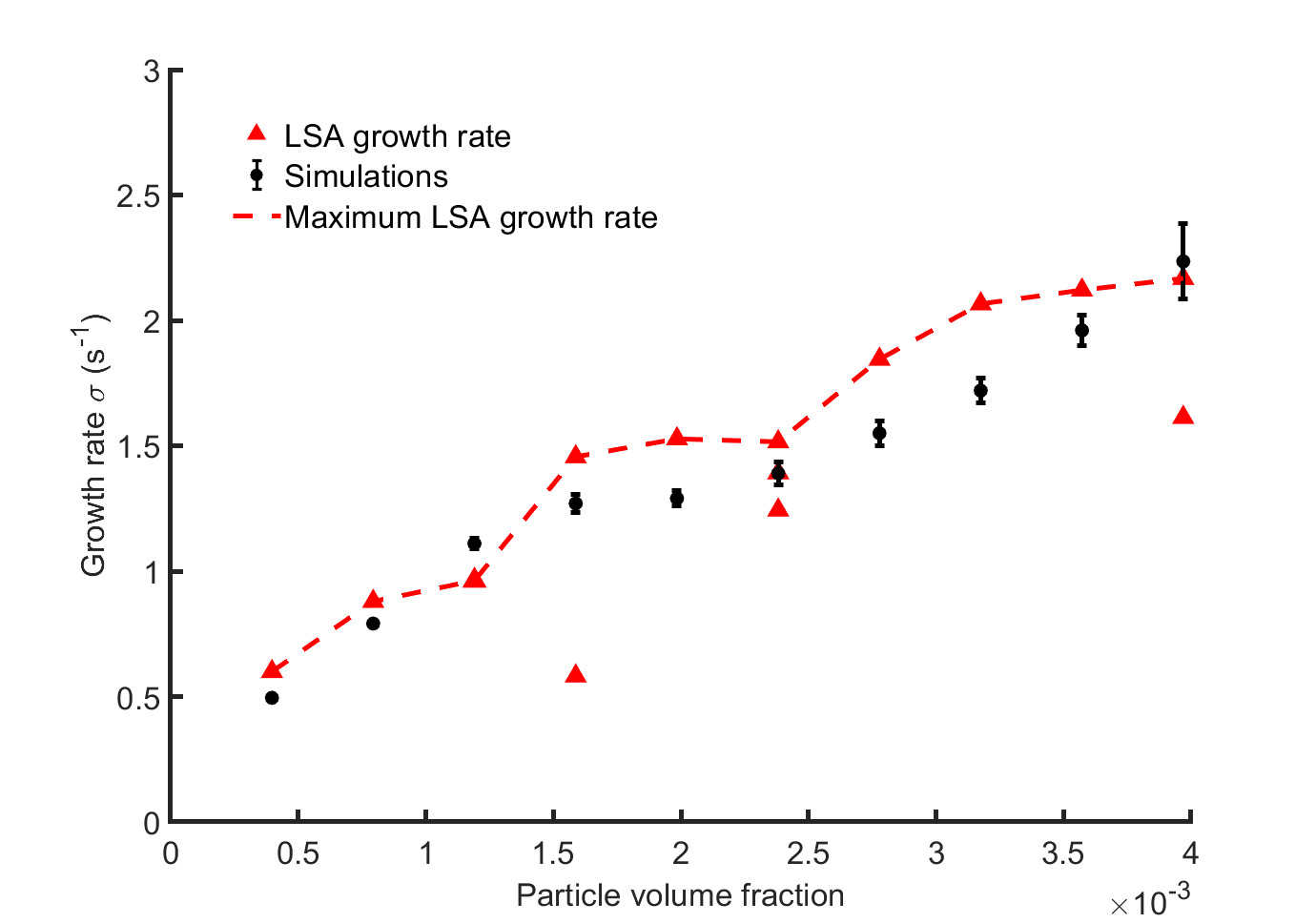}
\caption{Comparison of the instability growth rate measured in the simulations (black circles) and that predicted by the linear stability analysis (red triangles).}
\label{fig8}
\end{figure}

\subsection{Comparison with experimental investigations}

\begin{figure}[h]
\centering
\includegraphics[scale=0.9]{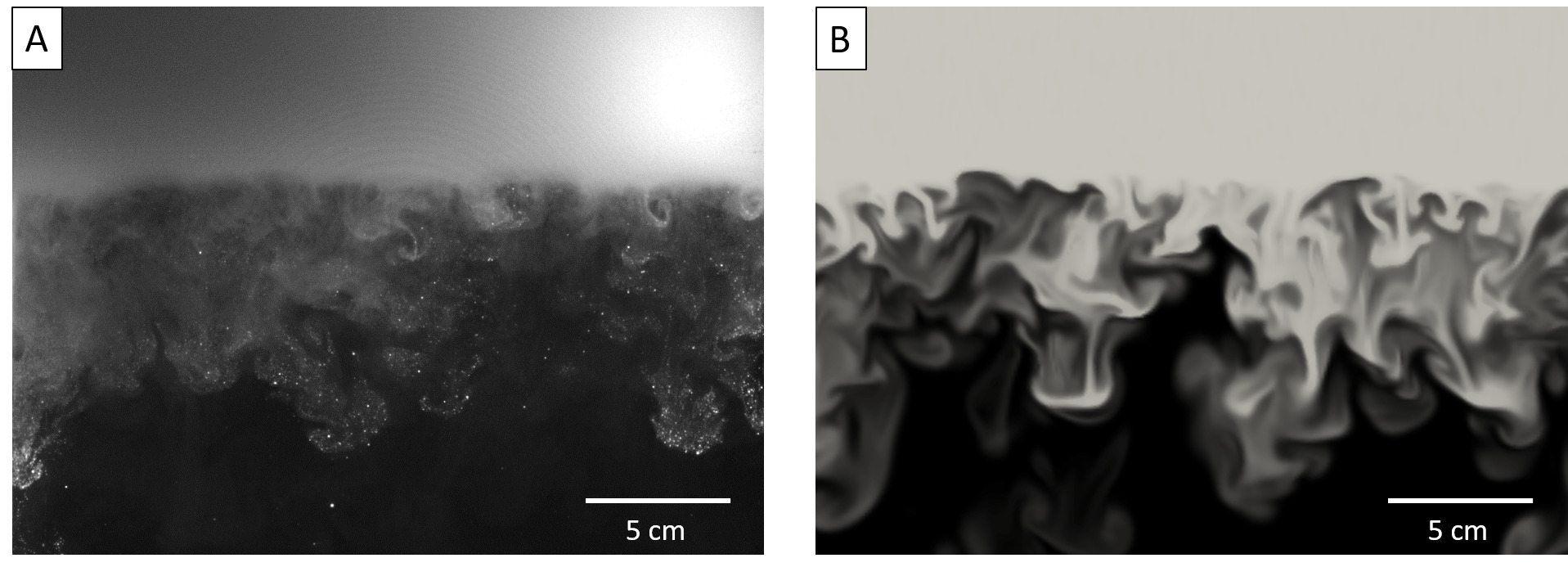}
\caption{Settling-driven gravitational instabilities observed 19.5 s after the barrier removal (A) in the laboratory \citep{Fries2021} and (B) in numerical simulations. Particle size: $40 \, \mu m$ and initial volume fraction : $\phi_0 = 2.78 \times 10^{-3}$.}
\label{fig9}
\end{figure}

Figures \ref{fig9}A and \ref{fig9}B show a qualitative comparison between snapshots taken from experiments \citep{Fries2021} and simulations (slice in the numerical domain). First, we note that our model is able to qualitatively reproduce the shape and size of fingers, especially their fronts where we observe the formation of lobes and eddies due to the Kelvin-Helmholtz instability \citep{Chou2016}. Second, we provide a quantitative validation of the non-linear regime by comparing our model with experiments, through measurements of the PBL thickness and the vertical finger velocity as functions of the particle volume fraction and size.

\subsubsection{Characterisation of the PBL and effect of the initial particle volume fraction on the finger velocity}

The bulk density profile $\rho_{blk}$, derived from the contributions of the particle concentration and sugar profiles, is given everywhere by the relation

\begin{equation}
\rho_{blk} = \phi \rho_p + \left( 1 - \phi \right) \rho_f.
\label{bulk}
\end{equation}

\begin{figure}[h]
\centering
\includegraphics[scale=0.9]{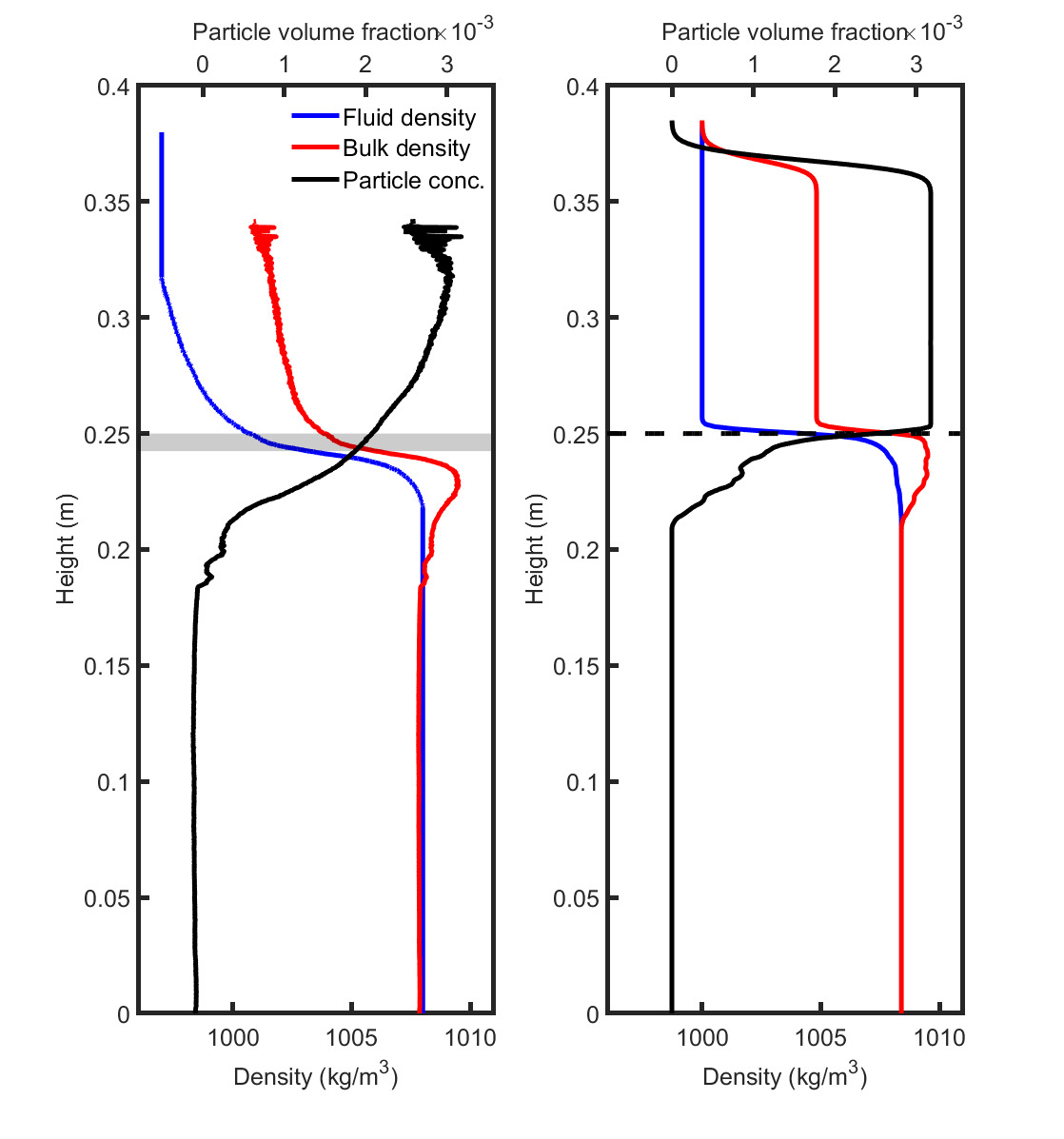}
\caption{Density profile after 8 seconds for experiments (left) \citep{Fries2021} and simulations (right) with $\phi_0 = 3.18 \times 10^{-3}$ and a particle size of $40 \, \mu m$.}
\label{fig10}
\end{figure}

\begin{figure}[h]
\centering
\includegraphics[scale=1.0]{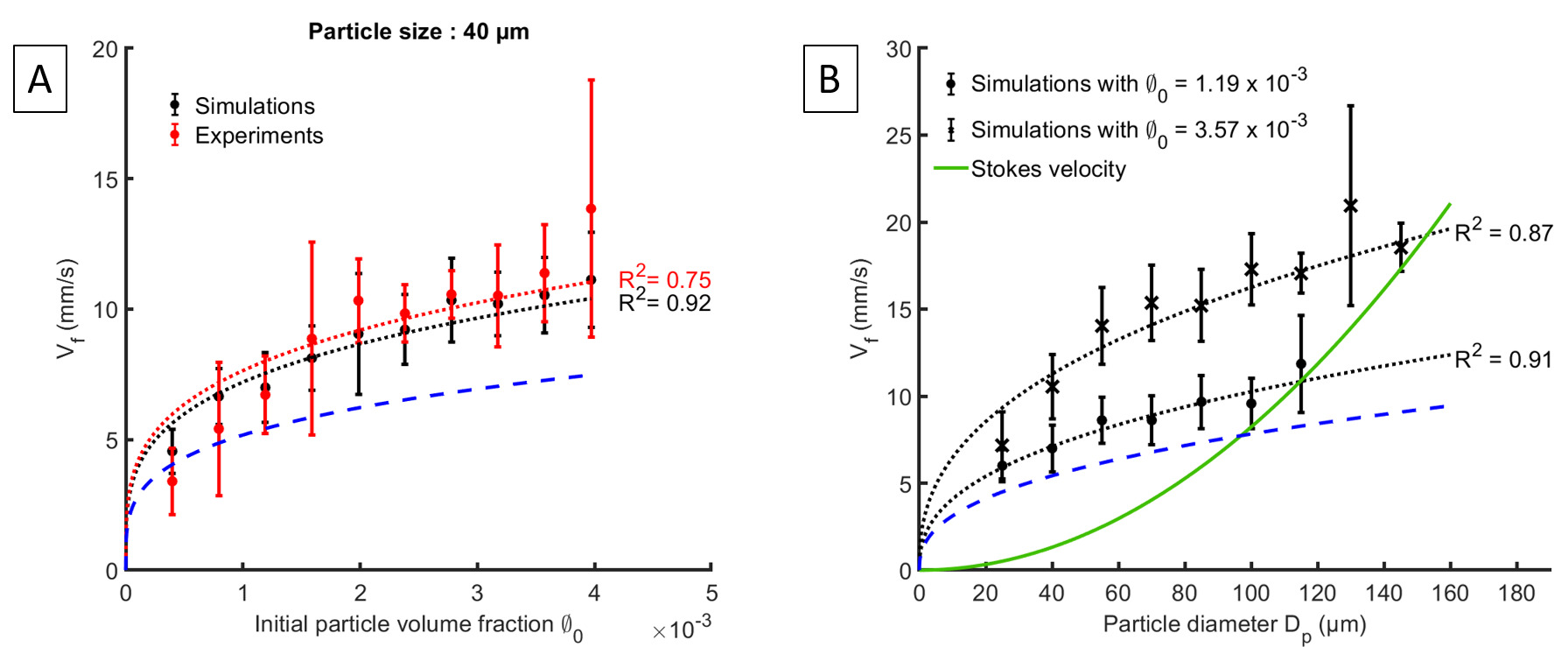}
\caption{(A) Average finger speed ($V_f$) as a function of the initial volume fraction ($\phi_0$) for a particle diameter of $40 \, \mu m$. Red and black dotted lines show the best fits to the experiments \citep{Fries2021} and simulations, respectively, using equation \ref{fing_vel} with $Gr_c$ as the fit parameter. For the simulations, we find $Gr_c = 1.2 \pm 0.4 \times 10^4$ whilst for the experiments $Gr_c = 1.9 \pm 0.7 \times 10^4$. (B) Average finger speed ($V_f$) as a function of the initial particle diameter ($D_p$), for two different particle volume fractions. The green line is the Stokes velocity for individual particles. The black dotted lines show the best fits to the simulations using equation \ref{fing_vel} with $Gr_c$ as the fit parameter. For $\phi_0 = 1.19 \times 10^{-3}$, the best fit gives $Gr_c = 7.6 \times 10^3$ and no fingers are observed to form for particle sizes higher than $115 \, \mu m$. For $\phi_0 = 3.57 \times 10^{-3}$, the best fit gives $Gr_c = 2.7 \times 10^4$ and no fingers are observed to form for particle sizes higher than $145 \, \mu m$. In the two plots, the blue dashed line shows equation \ref{fing_vel} using $Gr_c = 10^3$ from the analogy with thermal convection \citep{Hoyal1999}.}
\label{fig11}
\end{figure}

Figure \ref{fig10} shows the profiles of $\phi$, $\rho_f$ and $\rho_{blk}$ in the numerical simulations as well as in the experiments 8 seconds after the barrier removal for the same initial conditions ($\phi_0 = 3.18 \times 10^{-3}$ ). It can be seen that, in both the model and the experiments, there is an increase of the bulk density below the initial interface, owing to the particle front moving downwards. This zone of excess density corresponds to the unstable PBL from which instabilities occur, generating fingers. To calculate the finger velocity using the same method as in experiments, we extract slices from the 3D numerical domain and manually track the fronts of several fingers (6 to 15 fingers) from when they become fully developed until just before they become too diluted. For each simulation with different volume fraction, we then average the velocity of all tracked fingers and the uncertainty is the standard deviation associated with each set of fingers used for the measurements. Figure \ref{fig11}A shows the average finger velocity $V_f$ as a function of $\phi_0$, for both experiments \citep{Fries2021} and simulations. Our simulation results are in good agreement with the experimental measurements and highlight that the increase of $V_f$ with $\phi_0$ is non-linear.

By analogy with thermal convection, it has previously been assumed that $Gr_c = 10^3$ \citep{Hoyal1999}, but this is only an order of magnitude estimate and its application to settling-driven gravitational instabilities remains uncertain \citep{Fries2021}. Figure \ref{fig11}A shows good agreement between the simulations and equation \ref{fing_vel} for a fitted $Gr_c$ of $1.2 \pm 0.4 \times 10^4$ ($R^2=0.92$), which is an order of magnitude higher than the value previously assumed \citep{Carazzo2012, Hoyal1999}. This agrees reasonably with the experiments, where the best fit is obtained for $Gr_c = 1.9 \pm 0.7 \times 10^4$ ($R^2=0.75$), but the experimental results show more scatter. However, neither of these fits have completely satisfactory values of $R^2$. We therefore further investigate the applicability of equation \ref{fing_vel} by examining the dependence of $V_f$ on $\phi_0$, assuming a power law of the form $V_f \propto \phi_0^q$. According to equation \ref{fing_vel}, $q=4/15 \approx 0.27$. However, from the experiments, we obtain $q=0.50 \pm 0.16$ (with $R^2=0.95$) while for our simulations $q=0.37 \pm 0.08$ (with $R^2=0.98$). Here $Gr_c$, $q$ and their associated uncertainties have been calculated accounting for the uncertainty on $V_f$ with the SciPy (Python-based ecosystem) procedure $scipy.optimize.curve_fit$.

\subsubsection{Effect of particle size on the finger vertical velocity}

Since gravitational instabilities cause particles to sediment faster than their settling velocity, it is of interest to explore the transition from collective to individual settling, since this has implications for which grain sizes may prematurely sediment from a volcanic cloud \citep{Scollo2017}. Figure \ref{fig11}B shows the effect of particle size on the finger velocity as measured from the model, for two different initial volume fractions, in the experiments configuration (i.e. in the tank filled with water). We clearly observe two regimes:

\begin{itemize}
\item For particle sizes less than or equal to $115 \, \mu m$ (for $\phi_0 = 1.19 \times 10^{-3}$) and $145 \, \mu m$ (for $\phi_0 = 3.57 \times 10^{-3}$), we observe fingers, with the finger velocity increasing with particle size.
\item For greater particle sizes, no fingers are observed to form.
\end{itemize}

From our simulations, we constrain the transition between the two regimes to occur at a critical particle diameters around $115 \, \mu m$ and $145 \, \mu m$ respectively for $\phi_0 = 1.19 \times 10^{-3}$ and $\phi_0 = 3.57 \times 10^{-3}$. We also note that this size range corresponds to the particle size at which the Stokes velocity exceeds the predicted finger velocity. This result agrees with the experimental observations of \cite{Scollo2017}, who observed that no fingers form for particles with diameter larger than $\sim 125 \, \mu m$ with in initial particle volume fraction of $\phi_0 = 1.19 \time 10^{-3}$. We also compare the dependence of $V_f$ on the particle diameter with that predicted by equation \ref{fing_vel} and find a best fit for $Gr_c = 7.6 \pm 3.6 \times 10^3$ (with $R^2=0.91$) and $Gr_c = 2.7 \pm 0.8 \times 10^4$ (with $R^2=0.87$) respectively for the two initial volume fractions (Figure \ref{fig11}B). We observe again that the values for the fitted $Gr_c$ are greater than the one proposed by \cite{Hoyal1999} by analogy with thermal convection, whilst they also substantially differ from one another. We therefore also fit the results to a power law $V_f \propto D_p^ \eta $ finding $\eta = 0.38 \pm 0.13$ ($R^2=0.94$) and $\eta = 0.42 \pm 0.10$ ($R^2=0.88$) respectively to the two volume fractions which is in very good agreement with the analytical formulation (equation \ref{fing_vel}) that suggests $\eta = 0.4$.

\subsubsection{Particle mass flux, particle concentration in the lower layer and accumulation rate}

Given the excellent agreement between the proposed model and both LSA analysis and analogue experiments described above, we take advantage of having 3D data from the numerical simulations in order to extract other parameters which are difficult to obtain otherwise \citep{Fries2021}. Three interesting parameters are the particle mass flux across a plane, the particle concentration in the lower layer and the amount of particles accumulated at the bottom of the tank, which can be related to the accumulation rate. The latter is especially interesting as, when the model is applied to volcanic clouds, it could eventually be compared with field data \citep{Bonadonna2011a}.

\begin{figure}[h]
\centering
\includegraphics[scale=1.0]{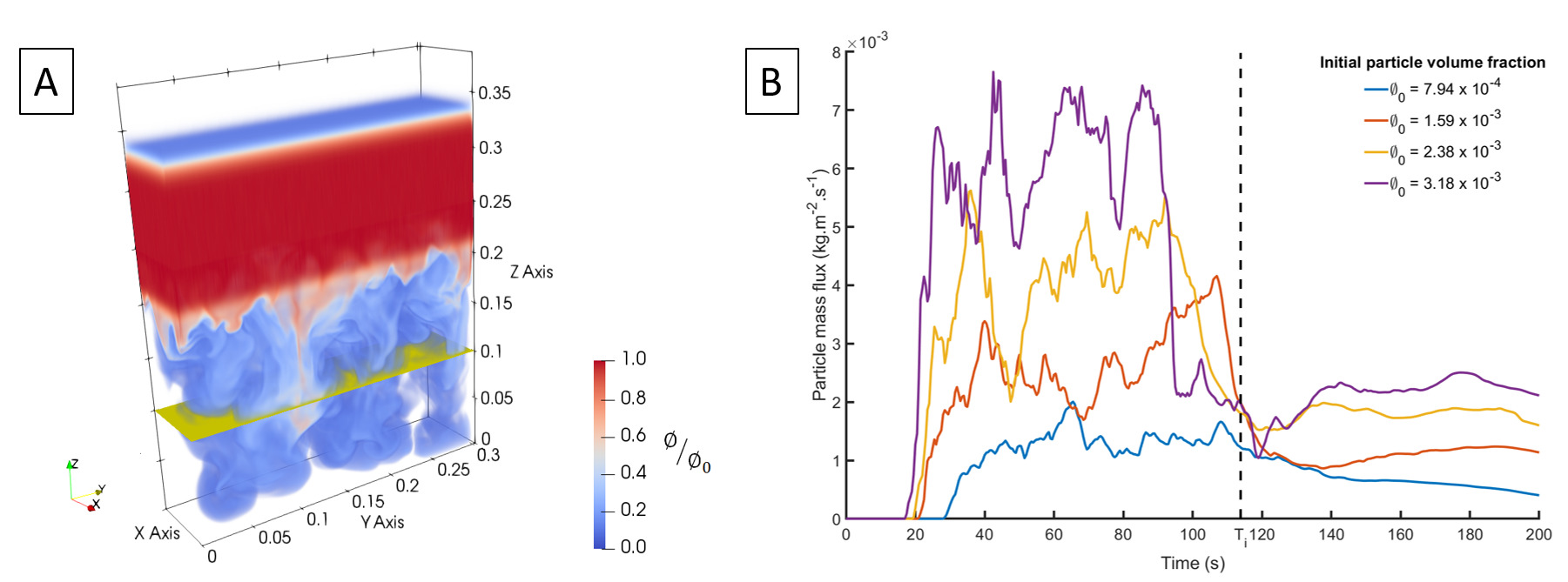}
\caption{(A) Horizontal planar surface (yellow slice) located $0.15 \, m$ below the barrier, across which the particle mass is computed in the simulation domain. (B) Temporal evolution for the mass of particles crossing the plane. Black dashed line: theoretical time for the particles to reach the plane at their individual Stokes velocity.}
\label{fig12}
\end{figure}

We calculate the mass flux across a horizontal plane (actually a thin box of thickness $\delta x$) as shown in Figure \ref{fig12}A  with

\begin{equation}
J = \frac{\Delta m}{A \Delta t},
\label{flux}
\end{equation}
where $\Delta m$ is the mass crossing the yellow plane of area $A$ in time $\Delta t$, and is given by the mass difference in the volume below the plane between $t$ and $t - \Delta t$. The mass below at each time is calculated by summing the mass of particles in each cell $i$ of volume $\Delta V$, which is individually given by $m_i = \Delta V \phi_i \rho_p$. Figure \ref{fig12}B shows the temporal evolution of the particle mass flux settling through the yellow plane (located at $0.15 \, m$ below the barrier), for several initial particle volume fractions. The vertical black dashed line indicates the theoretical time $T_i$ when particles would be expected to reach the plane if they were settling individually at their Stokes velocity. For the different simulations, we clearly observe that the moments when the flux starts initially increasing (i.e. the arrival of the fastest finger) are much earlier than $T_i$ and this shows the extent to which the collective settling enhances the premature sedimentation. After the initial increase, the fluxes exhibit strong oscillations around a high plateau. These oscillations are associated with the intermittent nature of PBL detachment and convection in the lower layer. Finally, the particle mass flux reaches a plateau after some time which shows the end of convection and a transition to individual settling. Throughout, the average mass flux, as well as the amplitude of the oscillations increases with the initial volume fraction.

Another way to highlight the enhancement of the sedimentation rate by collective settling is to study the spatial distribution of particles beneath the interface. Assuming a quiescent upper layer and a convective lower layer, akin to our simulations, \cite{Hoyal1999} derived equation \ref{C2_Hoy} for the evolution of the particle concentration in the lower layer.  The derivation of this formulation assumes that $\dot{M}_{out} \neq 0$ since $t=0$ but in fact, $\dot{M}_{out} = 0$ for $t < t_a$ where $t_a$ is the time when the first particles reach the bottom of the tank. Also, equation \ref{C2_Hoy} only remains valid  for $t<t_{lim}$, where $t_{lim}=h_1/V_s$ , $h_1$ being the thickness of the upper layer. After this time, there are no longer any particles remaining in the upper layer and $\dot{M}_{in} = 0$. We therefore propose an extension for the solution of the problem (see section 2.1 in Supplementary material) which becomes

\begin{equation}
\begin{array}{rcl} C_2 \left( t \right) = \frac{V_s}{h_2} C_1 \left( 0 \right) t, &  \mbox{for} \, & t<t_a, \end{array}
\label{C2_1}
\end{equation}

\begin{equation}
\begin{array}{rcl} C_2 \left( t \right) = C_1\left( 0 \right) \left[ 1 + \left( \frac{V_s}{h_2}t_a - 1 \right) e^{- \frac{V_s}{h_2} \left( t - t_a \right)} \right], &  \mbox{for} \, & t_a \leq t < t_{lim}, \end{array}
\label{C2_2}
\end{equation}

\begin{equation}
\begin{array}{rcl} C_2 \left( t \right) = C_1\left( 0 \right) \left[ 1 + \left( \frac{V_s}{h_2}t_a - 1 \right) e^{- \frac{V_s}{h_2} \left( t_{lim} - t_a \right)} \right] \left(1 + \frac{h_1}{h_2} - \frac{V_s}{h_2}t \right), &  \mbox{for} \, & t \geq  t_{lim}, \end{array}
\label{C2_3}
\end{equation}

where $h_2$ is the thickness of the lower layer. Equation \ref{C2_3} assumes that the convection stops at $t_{lim}$, which suggests a quiescent settling in the lower layer after that time with a constant flux $\dot{M}_{out}=AV_s C_2 \left( t_lim \right)$.

\begin{figure}[h]
\centering
\includegraphics[scale=1.0]{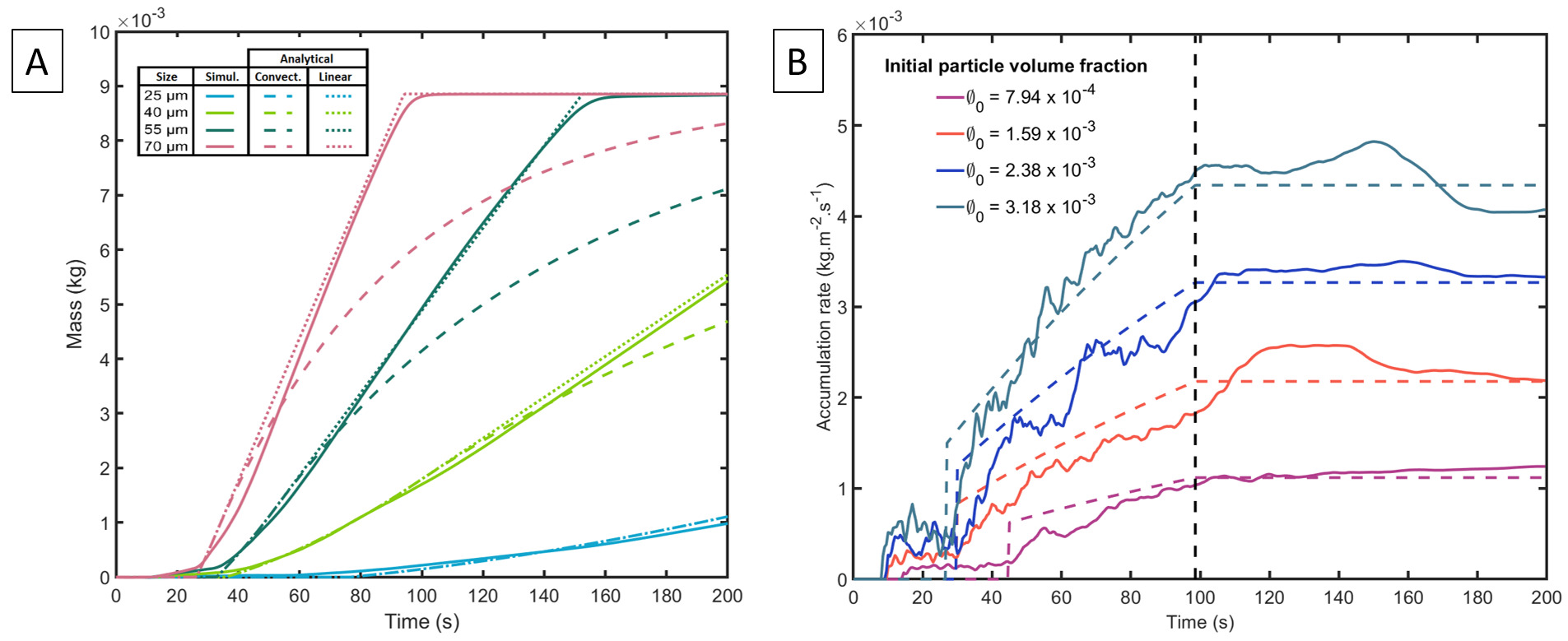}
\caption{(A) Temporal evolution of the mass of particles accumulating at the bottom of the tank for several particle sizes. The dashed and dotted lines represent the extended analytical model of \cite{Hoyal1999}. Particle volume fraction $\phi_0 = 1.19 \times 10^{-3}$. (B) Accumulation rate calculated at the bottom of the tank for several particle volume fractions and a particle size of $40 \, \mu m$. The coloured dashed lines are the rate derived from the analytical model. The black dashed line is the theoretical time at which all particles have settled across the interface.}
\label{fig13}
\end{figure}

An interesting result coming out of the previous analytical study is the mass of particles accumulating at the bottom of the tank and the associated accumulation rate. We can derive an analytical prediction for the mass of particles $m_b$ accumulated at the bottom of the tank for the different regimes highlighted above. Thus, by integration of the flux (see Section 2.2 in Supplementary material) we have

\begin{equation}
\begin{array}{rcl} m_b = 0, &  \mbox{for} \, & t<t_a, \end{array}
\label{m_1}
\end{equation}

\begin{equation}
\begin{array}{rcl} m_b = m_0 \frac{V_s}{h_1} \left[t + \left(\frac{h_2}{V_s} - t_a \right) e^{- \frac{V_s}{h_2} \left( t - t_a \right)} - \frac{h_2}{V_s} \right], &  \mbox{for} \, & t_a \leq t < t_{lim}, \end{array}
\label{m_2}
\end{equation}

\begin{equation}
\begin{array}{rcl} m_b = m_0 \frac{V_s}{h_1} \{ t_{lim} - \frac{h_2}{V_s} \left[ 1 + \left(\frac{V_s}{h_2} t_a - 1 \right) e^{- \frac{V_s}{h_2} \left( t_{lim} - t_a \right)} \right] \left(1 + \frac{h_1}{h_2} - \frac{V_s}{h_2} t \right) \}, &  \mbox{for} \, & t \geq  t_{lim}, \end{array}
\label{m_3}
\end{equation}
where $m_0$ is the initial mass of particles injected in the upper layer. Finally, at the time $t_{lim} + h_2/V_s$ , all the particle have settled through the lower layer, thus $m_b=m_0$. Figure \ref{fig13}A shows the simulated particle accumulation at the bottom of the tank through time, for different particle sizes as well as the analytical prediction (equations \ref{m_1}-\ref{m_3}). We compare as well with the analytical formulation of the mass which assumes that the lower is still turbulently convective even after the time $t_{lim}$ (equation S43 in Supplementary material, dashed lines in Figure \ref{fig13}A). In order to compare between this prediction and the model results, $t_a$ is fitted in order to have the best agreement between the numerical data and equations \ref{m_1}-\ref{m_3}. The results show clearly that the quiescent model of the lower layer for $t \geq t_{lim}$ agrees very well with the simulations and suggest that the entirely convective model underestimates the accumulation rate. Additionally, the fitted parameter $t_a$ is coherent with the time for the first fingers to reach the bottom of the tank in the simulations. Figure \ref{fig13}B shows the instantaneous accumulation rate computed from the numerical data for several initial volume fractions, as estimated by

\begin{equation}
\frac{1}{A} \frac{d m_b}{dt}.
\label{a_rate}
\end{equation}

\begin{figure}[h]
\centering
\includegraphics[scale=1.0]{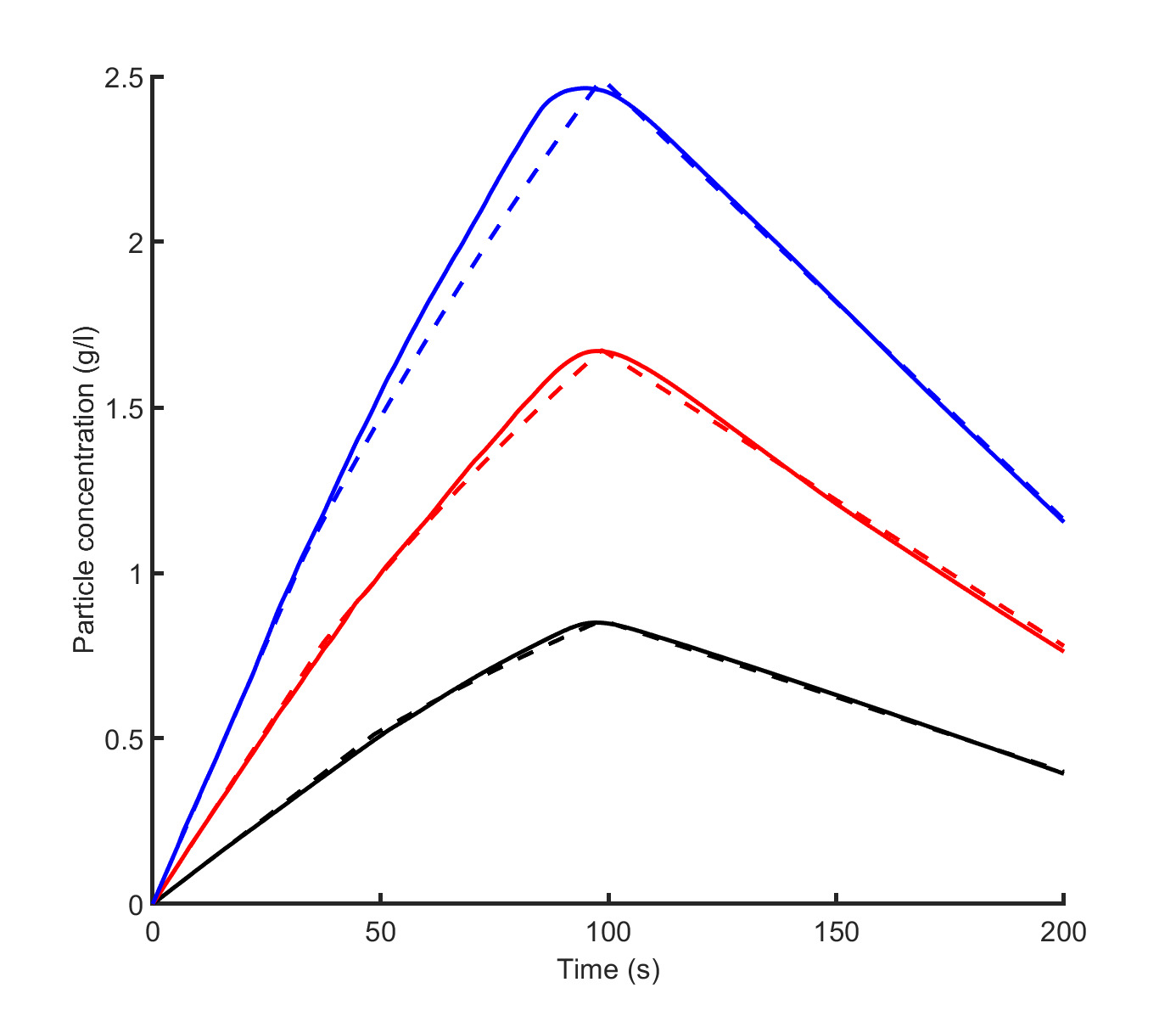}
\caption{Evolution of the average particle volume fraction in the lower layer for particle of size $40 \, \mu m$. Black: $C_0= 2 g/l$ $(\phi_0 = 7.94 \times 10^{-4})$, Red: $C_0= 4 g/l$ $(\phi_0 =1.59 \times 10^{-3})$ and Blue: $C_0= 6 g/l$ $(\phi_0 = 2.38 \times 10^{-3})$. Solid lines: numerical model. Dashed lines: modified \cite{Hoyal1999} model (equations \ref{C2_Hoy} and \ref{C2_1}-\ref{C2_3}).}
\label{fig14}
\end{figure}

We observe, for each initial particle volume fraction, an initial increase of the accumulation rate with time which reflects the enhancement of the sedimentation process due to convection. Interestingly, the accumulation rate then reaches a plateau at around $t=t_{lim}$, indicating that the system switches to a steady settling regime once all particles have left the upper layer. We compare also with the analytical relations which again have very good agreement with our simulations.
Finally, using the determined $t_a$, we can also calculate the concentration $C_2 \left( t \right)$, as calculated with the analytical expressions in equations \ref{C2_1}-\ref{C2_3}. Figure \ref{fig14} shows a comparison with the average $C_2 \left( t \right)$ as measured in simulations for a particle size of $40 \, \mu m$ and three different initial upper layer concentrations, finding very good agreement.

%------------------------------------------------

\section{Discussion}

\subsection{Model caveats}

Our numerical model has been validated by comparing various outputs with results from linear stability analysis, lab experiments \citep{Fries2021} and theoretical predictions from previous studies \citep{Carazzo2012, Hoyal1999}. Even though these comparisons are good (Figures \ref{fig9}-\ref{fig14}), the results provided by the model inherits the caveats of the experiments. Indeed, the static and confined configuration, as well as the fact that we performed the simulations in water, mean that we cannot fully extend the results to the volcanic case yet. Thus, further investigations are necessary to better simulate the volcanic environment (in air, presence of wind...). Additionally, it is necessary to consider the limits of validity of the assumption that the particles can be represented as a continuum. Whilst the condition on the particle coupling is given by the Stokes number ($St<1$), there is also a condition on the particle volume fraction to take in account. \cite{Harada2013} and \cite{Yamamoto2015} derived a dimensionless number in order to characterise the transition between fluid-like and particle-like settling. Although this number is only valid for narrow channel configurations, which are considerably different from ours, it highlights the fact that the particle size, volume fraction and characteristic length scale of the flow are critical parameters to define the validity of the continuum assumption. Thus, the transition from fluid-like to particle-like behaviour is achieved by decreasing the volume fraction and characteristic length scale and increasing the particle size. Near this transition, the use of a single-phase model, such as that presented here, should be treated with caution and this reveals the need for a comparison with future models which explicitly account for the drag contribution of individual particles.

Another related caveat concerns the numerical diffusion underlying the use of a Eulerian approach to describe the transport of particles. Compared to classical first order finite difference methods, the use of the third order WENO procedure has drastically reduced the numerical diffusion. It is also possible to further reduce the induced numerical diffusion by increasing the order of the WENO scheme (i.e. increase also the computational cost). However, for problems purely related to advection, where the presence of any diffusion is critical, another strategy, such as two-phase models (using a Lagrangian approach where individual particles are explicitly modelled), has to be considered. 

\subsection{Vertical finger velocity}

We have compared the simulated vertical velocity of fingers with experimental observations \citep{Fries2021} and a theoretical prediction (equation \ref{fing_vel}) from \citep{Carazzo2012, Hoyal1999} (Figure \ref{fig11}). This expression depends on a critical Grashof number, which by analogy with thermal convection \citep{Turner1973} has previously been assumed to be $10^3$ \citep{Hoyal1999}. This value effectively corresponds to a dimensionless critical PBL thickness at which point the PBL can detach and form fingers. However, both the model results and experimental observations summarised in Figure \ref{fig11} suggest that $Gr_c>10^3$ for our configuration. Furthermore, as seen in Figure \ref{fig11}B, the curve for $V_f$ using $Gr_c=10^3$  (blue dotted line) crosses the Stokes velocity curve around $95 \, \mu m$ for instance with an initial particle volume fraction of $\phi_0 = 1.19 \times 10^{-3}$, suggesting this value should be the upper particle size limit for finger formation. However, in agreement with experiments \citep{Fries2021, Scollo2017}, we observe a larger threshold for the finger formation to be in the size range $[115-125] \, \mu m$, for $\phi_0 = 1.19 \times 10^{-3}$, and in the range $[145-160] \, \mu m$ for $\phi_0 = 3.57 \times 10^{-3}$, in this particular configuration. We also showed that equation \ref{fing_vel} poorly predicts the observed dependence of the finger velocity on the initial particle volume fraction. Indeed, our studies suggests an alternative power law that better describes the dependence of $V_f$ on $\phi_0$. Equation \ref{fing_vel} has been derived by a scaling theory that involves $\delta_{PBL}$ as characteristic length of the problem \citep{Carazzo2012, Hoyal1999} and the discrepancies highlighted in this paper (Figure \ref{fig11}) may suggest that $\delta_{PBL}$ actually has a slightly different dependence on the initial particle volume fraction. Moreover, the use of the Grashof number as an appropriate scaling for the PBL thickness remains uncertain. On one hand, our results suggest that if instability does occur once a critical Grashof number is reached, the critical value taken from the thermal convection analogy is not valid. On the other hand, the Grashof number may simply not be the correct dimensionless form of the PBL thickness, and different flow configurations will produce different critical values. However, the predicted dependence of the finger velocity on the particle diameter by equation \ref{fing_vel} shows a very good agreement with our simulated results, as confirmed by a power-law fitting between $V_f$ and $D_p$. Thus, whilst we have demonstrated the need for a better scaling of $\delta_{PBL}$, equation \ref{fing_vel} can still provide a good estimate for the particle size threshold to form fingers. Consequently, if the size threshold to form fingers is given when equation \ref{fing_vel} equals the Stokes velocity (equation \ref{stokes_vel}) we can derive a formulation for the threshold

\begin{equation}
D^{*}_p = \left[ \frac{\left( 18 \mu \right)^2 \phi \delta_{PBL}}{g \left( \rho_p - \rho_f \right) \rho_f} \sqrt{\left( \frac{\pi}{4} \right)} \right] ^ {\frac{1}{4}},
\label{thres}
\end{equation}

The main caveats for this formulation are that it strongly depends on having a correct scaling for $\delta_{PBL}$ and obviously, this estimation is valid under the assumption that particles settle at their Stokes velocity, which is reasonable for our study but might be uncertain in nature where the ambient fluid is air and for non-spherical particles.

\subsection{Particle concentration in the lower layer and mass accumulation rate}

We have proposed a modified analytical formulation for the particle concentration in the lower layer $C_2 \left( t \right)$ and consequently for the mass of particles accumulated at the bottom of the tank $m_b \left( t \right)$. Despite some numerical artefacts that can be seen on Figure \ref{fig13}B where the computed accumulation rate seems to be non-zero before $t_a$, there is very good agreement between the simulations and the analytical model. The artefacts themselves are due to fluctuating numerical errors that do not affect the final results.

The analytical predictions for $C_2 \left( t \right)$ and $m_b \left( t \right)$ are step-wise functions depending on $t_a$, the time it takes for the first particles to reach the bottom. For $t<t_a$, the analytical model predicts that $C_2 \left( t \right)$ increases linearly with time since the formulation assumes that, during this period, particles are settling individually. In fact, our numerical results show that convective settling does occur for $t<t_a$ but, since this time period is short, the linear law seems to be a satisfactory approximation for the early-time average lower layer particle concentration. However, in order to compare our simulated results with the analytical prediction, we fitted the parameter $t_a$ in this study. Although we are able to obtain excellent agreement between model and theory, it would be better to develop a fully independent formulation. To achieve this, it is necessary to also provide an analytical estimation for $t_a$. One possible approach would be to assume the decomposition $t_a=t_a^{'}+t_a^{''}$ where $t_a^{'}$ is the time during which the PBL initially grows beneath the interface at the individual particle settling velocity, i.e., $t_a^{'}=\delta_{PBL}/V_s$ , and $t_a^{''}$ is the time between the PBL detachment and the first arrival of particles at the base of the domain. If, during this stage, we assume that the particles are advected at the finger velocity then $t_a^{''} = \left( h_2-\delta_{PBL} \right)/V_f$. We therefore see that $t_a$ strongly depends on $\delta_{PBL}$, which highlights once again the need for a correct scaling of the PBL thickness, as discussed in the previous section.
Another interesting result concerns the accumulation rate of particles at the base of the domain in the presence of fingers. Figure \ref{fig13}B shows the accumulation rate increases with time for $t_a<t<t_{lim}$, in agreement with the analytical prediction (i.e. combination of equations \ref{m_2} and \ref{a_rate} which provides an exponential increase of $m_b$). Conversely, if the particles had settled individually, the accumulation rate would be temporally constant. This shows that temporally resolved measurements of the accumulation rate of particles from volcanic clouds may record temporal signatures of sedimentation via settling-driven gravitational instabilities. Whilst there is already a spatial deposit signature of settling-driven gravitational instabilities (i.e. bimodal grainsize distribution) \citep{Bonadonna2011a, Manzella2015}, this is not unique and can be generated by other mechanisms such as particle aggregation \citep{Brown2012}. Accumulation rate data from the field may therefore provide a powerful tool for distinguishing the efficiency of convective sedimentation beneath volcanic clouds.

\section{Conclusions}

We have presented an innovative hybrid Lattice Boltzmann Finite Difference 3D model in order to simulate settling-driven gravitational instabilities at the base of volcanic ash clouds. Such instabilities occur when particles settle through a density interface at the base of a suspension, leading to the formation of an unstable particle boundary layer \citep{Carazzo2012, Hoyal1999, Manzella2015}, and also occur in other natural settings, such as river plumes \citep{DavarpanahJazi2016}. Our numerical model makes use of a low-diffusive WENO procedure to solve the advection-diffusion-settling equation for the particle volume fraction. The use of such a routine allows us to minimise errors associated with numerical diffusion and has the advantage of being applied to simple uniform meshes, which makes the coupling with the LBM easier. This innovative use of the WENO scheme, therefore, represents an effective tool for the solving of advection-dominated problems. Our implementation of the third order WENO finite difference scheme will be integrated in a future release of the open-source $Palabos$ code. Our model has been successfully validated by comparing the results with i) predictions from linear stability analysis where we show that the model is able to simulate settling-driven gravitational instabilities from the initial disturbance through the linearly-unstable regime, ii) analogue experiments \citep{Fries2021} and iii) theoretical models \citep{Carazzo2012, Hoyal1999} in order to reproduce the non-linear regime which describes the downward propagation of fingers. We also confirmed the premature sedimentation process through collective settling compared to individual settling.

Our model provides new insights into:

\begin{itemize}
\item the value of the critical Grashof number. From measurements of the vertical finger speed, we have found $Gr_c \sim 10^4$ in our configuration. This value differs from the one suggested by analogy with thermal convection ($Gr_c \sim 10^3$) \citep{Hoyal1999}. Our results suggest that either the critical Grashof number for settling-driven gravitational instabilities is greater than in the thermal convection case or that the Grashof number may not be the correct dimensionless form of the PBL thickness. In any case, this highlights the need for further investigation of the scaling of the PBL thickness $\delta_{PBL}$.
\item the presence of a particle size threshold for the finger formation. Using our results, we have proposed an analytical formulation for this threshold depending on the density of particles, the viscosity of the medium and also the bulk density difference between the two fluid layers.
\item the signature of settling-driven gravitational instabilities (i.e. accumulation rate). We show that the accumulation rate of particles at the tank base initially increases with time before reaching a plateau. This contrasts with the constant accumulation rate associated with individual particle settling. This suggests that accumulation rate data could be used during tephra fallout to distinguish between sedimentation through settling-driven gravitational instabilities and individual-particle sedimentation. 
\end{itemize}

We have also demonstrated how our numerical model can be used to expand the initial conditions and configuration settings that can be explored through experimental investigations. The results presented so far in an aqueous media permitted model validation but have also opened fundamental questions that will be addressed in future works involving configurations more similar to the natural system. Indeed, thanks to the strengths of the LBM, the model can easily be applied to more complex systems and provide a robust tool for the transition from the laboratory studies to volcanic systems, as well as other environmental flows.

\section*{Conflict of Interest}

The authors declare that the research was conducted in the absence of any commercial or financial relationships that could be construed as a potential conflict of interest.

\section*{Data Availability Statement}

The datasets generated for this study can be found in the article/Supplementary Material. Additional raw data are available upon request.

\section*{Author Contributions}

Jonathan Lemus integrated the WENO procedure in the Palabos framework, conducted the simulations and data analysis under the supervision of Paul Jarvis, Jonas Lätt, Costanza Bonadonna and Bastien Chopard. Jonathan Lemus drafted the manuscript. Jonas Lätt and Bastien Chopard were involved in the development of the Palabos code.  All authors have contributed to data interpretation as well as the editing and finalising of the paper.

\section*{Funding}

The study has been funded by the Swiss National Science Foundation $\sharp 200021 \_ 169463$.

\section*{Acknowledgments}

All the simulations presented in this paper have been performed using the High Performance Computing (HPC) facilities $Baobab$ and $Yggdrasil$ of the University of Geneva. We would like to thank Amanda B. Clarke and Jeremy C. Phillips for constructive discussions about the problem.

%----------------------------------------------------------------------------------------
%	REFERENCE LIST
%----------------------------------------------------------------------------------------

\bibliographystyle{apalike}
\bibliography{library}

%----------------------------------------------------------------------------------------

\end{document}

% Print the title
\maketitle

\section{Description of the finite difference schemes}

\subsection{First-order upwind finite difference scheme}

In the following description, we describe only the one-dimensional scheme as the method is easily generalised to higher dimensions by simply applying the procedure separately to each dimension \citep{Ferziger2002}. The finite difference method is based on the approximation of the derivatives at the node locations of a discretised domain and can be applied to both uniform and non-uniform meshes. However, we describe here the case where the numerical domain is uniformly discretised with the spatial step $\delta x$, such that the domain is divided in a set of equally-spaced points $\{x_0, x_1, ..., x_i ..., x_n \}$. We also consider the one-dimensional conservation equation

\begin{equation}
\frac{\partial a}{\partial t} + u \frac{\partial a}{\partial x} = 0,
\label{cons}
\tag{S1}
\end{equation}

where $a$ is the transported information and $u$ the advection velocity. In our discrete domain, we use the notation $a_i^n  =a \left(x = x_i, t = t_n \right)$,with $n$ denoting the time coordinate and $i$ the spatial coordinate.

Among the numerous methods used to approximate the derivative $\partial a / \partial x$ at location $x_i$, the Taylor expansion is the most common. Following this procedure, there are two ways to estimate the derivative with a first order accuracy:

\begin{itemize}
\item Using the Taylor expansion at the location $x_{i+1} = x_i + \delta x$, we get the forward difference

\begin{equation}
\left( \frac{\partial a}{\partial x} \right)_F \approx \frac{a^n_{i+1} - a^n_i}{\delta x}.
\label{forw}
\tag{S2}
\end{equation}

\item Using the Taylor expansion at the location $x_{i-1} = x_i - \delta x$ we get the backward difference

\begin{equation}
\left( \frac{\partial a}{\partial x} \right)_B \approx \frac{a^n_{i} - a^n_{i-1}}{\delta x}.
\label{backw}
\tag{S3}
\end{equation}

\end{itemize}

The central finite difference approximation can be determined by combining the forward and backward differences. Thus, we have the second order accurate discretisation

\begin{equation}
\left( \frac{\partial a}{\partial x} \right)_C \approx \frac{a^n_{i+1} - a^n_{i-1}}{2 \delta x}.
\label{cent}
\tag{S4}
\end{equation}

The upwind finite difference method scheme is an adaptive procedure to discretise the problem based on the direction of propagation of the information. The estimation of the quantity $a_i^{n+1}$ depends on the sign of $u$:

\begin{itemize}
\item If $u>0$, the backward difference is used
\begin{equation}
a_i^{n+1} = a_i^{n} - \frac{\delta t}{\delta x} u \left[a_i^{n} - a_{i-1}^{n} \right].
\label{backw2}
\tag{S5}
\end{equation}

\item If $u<0$, the forward difference is used

\begin{equation}
a_i^{n+1} = a_i^{n} - \frac{\delta t}{\delta x} u \left[a_{i+1}^{n} - a_{i}^{n} \right].
\label{forw2}
\tag{S6}
\end{equation}

\end{itemize}

All these properties of the first order upwind scheme guarantee a strong stability of the solution providing the Courant–Friedrichs–Lewy (CFL) condition is satisfied  \citep{Courant1928}

\begin{equation}
\frac{\delta t}{\delta x} u \leq 1.
\label{CFL}
\tag{S7}
\end{equation}

\subsection{Third-order Weighted Essentially Non-Oscillatory (WENO) finite-difference scheme}

\begin{figure}[h]
\centering
\includegraphics[scale=0.8]{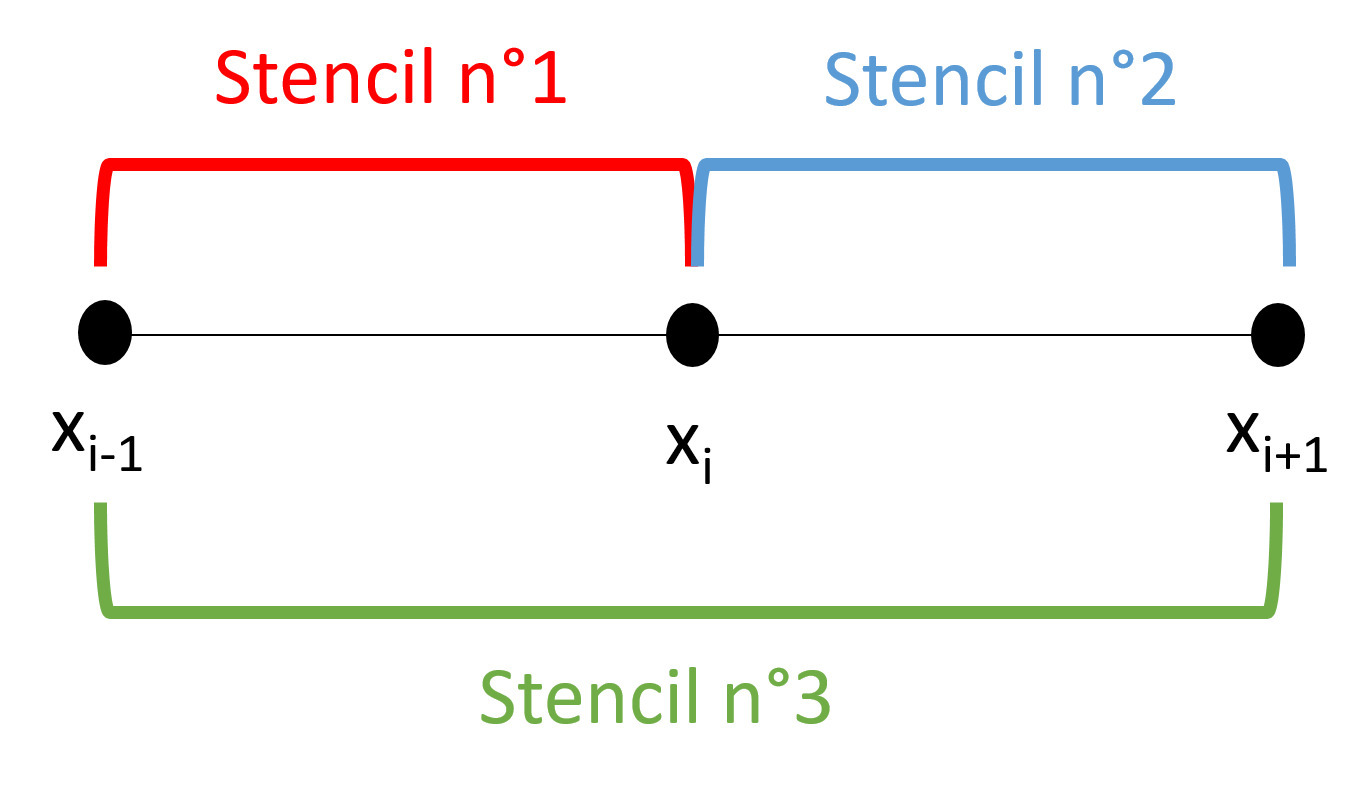}
\caption{Numerical stencil used for the 3rd order WENO procedure}
\label{S1}
\end{figure}

In the following description, the numerical domain is discretized using the same set of points as presented in the previous section. The low-diffusive WENO scheme belongs to a family of high-resolution methods and was developed in order to solve hyperbolic partial differential equations of the form

\begin{equation}
\frac{\partial a}{\partial t} + \frac{\partial h}{\partial x} = 0,
\label{PDE}
\tag{S8}
\end{equation}
where, in our case, the flux takes the form  $ h \left( a \right) = u a \left( x \right)$ \citep{Jiang1996, Liu1994}. The WENO scheme provides a third order accurate method in smooth regions, i.e., where the spatial gradient is small. On the other hand, the scheme adapts where the gradient is high. In these cases, the accuracy tends toward second-order. This adaptive aspect of the WENO procedure ensures the suppression of spurious oscillations, predicted by the Godunov theorem \citep{godunov1954different, godunov1959difference}, around shocks (regions of high gradient). The principle is to build a convex combination of interpolants for the flux at given points of the domain, using different stencils. The third-order WENO procedure uses two adjacent stencils of two points each (Figure S\ref{S1}).

It is possible to approximate the spatial derivative with the commonly used half-node flux:

\begin{equation}
\frac{\partial h}{\partial x} \approx \frac{h_{i+\frac{1}{2}}-h_{i-\frac{1}{2}}}{\delta x},
\label{approx}
\tag{S9}
\end{equation}
where $h_{i+\frac{1}{2}} = h \left( a \left( x_{i+\frac{1}{2}} \right) \right)$ is the numerical flux at the half-node location $ x_{i+\frac{1}{2}} $ (i.e., the central position between the points $x_i$ and $x_{i+1}$). There are three ways to construct the half node flux $h_{i+\frac{1}{2}}$. Firstly, we can use polynomial interpolants of degree 1 $ p_1\left( x \right) $ and $ p_2\left( x \right) $, respectively, in the two-point stencils 1 and 2. Then, the flux at the location $ x_{i+\frac{1}{2}} $ is given either by $h^{(1)}_{i+\frac{1}{2}} \equiv  p_1\left( x_{i+\frac{1}{2}} \right)$ or $h^{(2)}_{i+\frac{1}{2}} \equiv  p_2\left( x_{i+\frac{1}{2}} \right)$ and for $u>0$ we have

\begin{equation}
h^{(1)}_{i+\frac{1}{2}} = \frac{1}{2} \left( h_{i} + h_{i+1} \right),
\label{Weno1}
\tag{S10}
\end{equation}
and

\begin{equation}
h^{(2)}_{i+\frac{1}{2}} = - \frac{1}{2} \left( h_{i-1} - 3 h_{i} \right).
\label{Weno2}
\tag{S11}
\end{equation}

Note that the flux approximations for each stencil in this case have an accuracy of second order. The last way to determine the flux $h_{i+\frac{1}{2}}$ is with an interpolating polynomial $p_3 \left( x \right)$ of degree 2 inside the three-point stencil 3 (i.e., the union of stencils 1 and 2 in Figure S\ref{S1}). Then, the third-order flux at the location $x_{i+\frac{1}{2}}$ is given by

\begin{equation}
h^{(3)}_{i+\frac{1}{2}} =\frac{3}{4} h_i - \frac{1}{8} h_{i-1} + \frac{3}{8} h_{i+1}.
\label{Weno3}
\tag{S12}
\end{equation}

The three interpolations of the half-node flux presented above are efficient assuming that the function h is smooth through the associated stencil and it is even possible to write the third-order flux $ h^{(3)}_{i+\frac{1}{2}} $ as a linear combination of $ h^{(1)}_{i+\frac{1}{2}} $ and $ h^{(2)}_{i+\frac{1}{2}} $

\begin{equation}
h^{(3)}_{i+\frac{1}{2}} = \gamma_1 h^{(1)}_{i+\frac{1}{2}} + \gamma_2 h^{(2)}_{i+\frac{1}{2}},
\label{Weno4}
\tag{S13}
\end{equation}
where $\gamma_1=3/4$ and $\gamma_2=1/4$. However, the presence of any discontinuity would break the stability of the procedure, introducing spurious oscillations in the calculated solution. The treatment of this aspect constitutes the essence of the WENO method which retains the property of relating the total flux $h_{i+\frac{1}{2}}$ to a convex combination $ h^{(1)}_{i+\frac{1}{2}} $ and $ h^{(2)}_{i+\frac{1}{2}} $, similarly to equation \ref{Weno4}, but including non-linear weights

\begin{equation}
h^{(3)}_{i+\frac{1}{2}} = \omega_1 h^{(1)}_{i+\frac{1}{2}} + \omega_2 h^{(2)}_{i+\frac{1}{2}},
\label{Weno5}
\tag{S14}
\end{equation}
where $\omega_1$ and $\omega_2$ are functions of the smoothness of $h$ and must satisfy $\omega_1+\omega_2 = 1$. If $h$ is smooth in all the stencils, $\omega_i \rightarrow \gamma_i$ and ensures a third-order accuracy. Conversely, the presence of any discontinuity in the $i-th$ stencil means $\omega_i \rightarrow 0$, decreasing the accuracy to second-order. This property is guaranteed by determining $\omega_i$

\begin{equation}
\omega_i = \frac{\tilde{\omega}_i}{\sum_j \tilde{\omega}_j},
\label{omega}
\tag{S15}
\end{equation}
where
\begin{equation}
\tilde{\omega_j} = \frac{\alpha_j}{\left( \epsilon + \beta_j \right)^2},
\label{omega}
\tag{S16}
\end{equation}
and, in the third-order WENO case, $\alpha_1=  2/3$, $\alpha_2 = 1/3$, $\beta_1=\left( h_{i+1}-h_i \right)^2$ and $\beta_2 = \left( h_i - h_{i-1} \right)^2$ ($\beta_j$ are referred to as smoothness indicators). A positive mathematical coefficient $\epsilon$ is introduced in order to avoid any division by zero for the calculation of  $\tilde{\omega}_j$.

The WENO reconstruction presented above is valid for $u>0$. In order to include potential changes of the flow direction, we use an upwind splitting of the flux. Thus, as a mirror image of the procedure described above, we have for $u<0$

\begin{equation}
h^{(1)}_{i+\frac{1}{2}} = - \frac{1}{2} \left( h_{i+2} - 3 h_{i+1} \right),
\label{Weno1bis}
\tag{S17}
\end{equation}
and

\begin{equation}
h^{(2)}_{i+\frac{1}{2}} = \frac{1}{2} \left( h_{i+1} + h_{i} \right).
\label{Weno2bis}
\tag{S18}
\end{equation}

\begin{equation}
\beta_1=\left( h_{i+1}-h_{i+2}\right)^2,
\label{beta1}
\tag{S19}
\end{equation}
and

\begin{equation}
\beta_2=\left( h_{i}-h_{i+1}\right)^2.
\label{beta1}
\tag{S20}
\end{equation}

With an appropriate time discretisation, the third-order WENO procedure remains a Total Variation Diminishing (TVD) scheme, i.e., $\sum_i | a^{n+1}_{i+1} - a^{n+1}_{i}| \leq \sum_i | a^{n}_{i+1} - a^{n}_{i}|$ \citep{Harten1983}. This property avoids the introduction of new local extrema in the solution. Therefore, a TVD third-order WENO scheme is given using an iterative third-order Runge-Kutta method for the time discretisation:

\begin{equation}
a^{(1)} = a^{n} + \delta t W\left( a^n \right),
\label{RK1}
\tag{S21}
\end{equation}

\begin{equation}
a^{(2)} = \frac{3}{4} a^{n} + \frac{1}{4} a^{(1)} + \frac{1}{4} \delta t W\left( a^{(1)} \right),
\label{RK2}
\tag{S22}
\end{equation}
and

\begin{equation}
a^{n+1} = \frac{1}{3} a^{n} + \frac{2}{3} a^{(2)} + \frac{2}{3} \delta t W\left( a^{(2)} \right),
\label{RK3}
\tag{S23}
\end{equation}

with $W(a)$ is the spatial derivative given by the WENO algorithm described above.

\section{Extension of analytical model of \cite{Hoyal1999}}

\subsection{Particle concentration}

For a particle suspension placed above a denser fluid, an estimate for the evolution of the particle concentration in the lower layer has been derived by \cite{Hoyal1999}. We provide here an extension of this formulation.

\subsubsection{Upper layer}

First, let’s consider a quiescent upper layer of thickness $h_1$ with an initial particle concentration $C_1 \left(0 \right)$. For a quiescent settling at the velocity $V_s$, the constant flux of particles leaving the upper layer is given by $-A V_s C_1 \left( 0 \right)$. Thus, the evolution for the mass of particles is described by the equation:

\begin{equation}
\frac{d M_1}{dt} = -A V_s C_1 \left( 0 \right).
\label{mass_bal1}
\tag{S24}
\end{equation}

Assuming that the mass of particles depends on the concentration as:

\begin{equation}
M_i \left( t \right) = A h_i C_i \left( t \right),
\label{mass_bal2}
\tag{S25}
\end{equation}
$A$ being the horizontal cross section of the tank (where $i=1$ for the upper layer and $i=2$ for the lower layer), the temporal evolution of the particle concentration in the upper layer is the solution of:

\begin{equation}
\frac{d C_1}{dt} = -\frac{V_s}{h_1} C_1 \left( 0 \right).
\label{mass_bal3}
\tag{S26}
\end{equation}
Then:

\begin{equation}
C_1 \left( t \right) = C_1 \left( 0 \right) \left[1 - \frac{V_s}{h_1} t \right].
\label{mass_bal4}
\tag{S27}
\end{equation}

\subsubsection{Lower layer}

Now, let’s define $\dot{M}_{in}$ as the flux of particle entering the lower layer (i.e. the particles arriving from the upper layer) and $\dot{M}_{out}$ the flux of particles leaving (i.e. particles that deposit at the bottom of the tank). Thus, the mass of particle in the lower layer $M_2 \left( t \right)$ is governed by the equation

\begin{equation}
\frac{d M_2}{dt} = \dot{M}_{in} - \dot{M}_{out}, 
\label{mass_bal5}
\tag{S28}
\end{equation}

\begin{itemize}
\item 	There is a time $t_a$ before which particles have not yet reached the bottom of the tank and evidently $ \dot{M}_{out} = 0$. Still assuming a quiescent upper layer, we have also:
\begin{equation}
\dot{M}_{in} = A V_s C_1 \left( 0 \right).
\label{mass_bal6}
\tag{S29}
\end{equation}

So, combining \ref{mass_bal2}, \ref{mass_bal5} and \ref{mass_bal6} the particle concentration in the lower layer for $t<t_a$ is the solution of :

\begin{equation}
\frac{d C_2}{dt} =\frac{V_s}{h_2} C_1 \left( 0 \right).
\label{mass_bal7}
\tag{S30}
\end{equation}

that is to say (assuming $C_2 \left( 0 \right) = 0$):

\begin{equation}
C_2 \left( t \right) = \frac{V_s}{h_2} C_1 \left( 0 \right) t .
\label{mass_bal8}
\tag{S31}
\end{equation}

\item Once the first particles have reached the bottom of the tank (i.e. $t \geq t_a$), the flux of particles leaving the lower layer becomes

\begin{equation}
\dot{M}_{out} = A V_s C_2 \left( t \right).
\label{mass_bal9}
\tag{S32}
\end{equation}

Similarly to the previous point, combining equations \ref{mass_bal2} (which assumes that the lower layer is turbulently convecting and is homogeneous), \ref{mass_bal5}, \ref{mass_bal6} and \ref{mass_bal8} we obtain

\begin{equation}
\frac{d C_2}{dt} + \frac{V_s}{h_2} C_2 =\frac{V_s}{h_2} C_1 \left( 0 \right).
\label{mass_bal10}
\tag{S33}
\end{equation}

Using the initial condition $C_2 \left( t_a \right) = (V_s/h_2 ) C_1 \left( 0 \right) t_a$ (continuity with \ref{mass_bal8}), the solution of this equation is:

\begin{equation}
C_2 \left( t \right) = C_1 \left( 0 \right) \left[ 1 + \left( \frac{V_s}{h_2} t_a - 1 \right) e^{- \frac{V_s}{h_2} \left( t - t_a \right)} \right].
\label{mass_bal11}
\tag{S34}
\end{equation}

It is evident that when $t_a = 0$, \ref{mass_bal11} is equivalent to the original formulation of \cite{Hoyal1999} (equation 4 in the Introduction).

\item However, \ref{mass_bal11} is only valid for the condition that particles keep settling across the interface i.e. for a time $t<t_{lim}$ (where $t_{lim}=h_1/V_s$ ). We have extended the formulation to later times once all particles from the upper layer have settled across the interface i.e. for a time $t \geq t_{lim}$. As there are no longer particles in the upper layer, the flux $\dot{M}_{in}$ drops to zero. Two possibilities are now available. On one hand, if we consider that the lower layer is still turbulently convecting after $t_{lim}$, equation \ref{mass_bal10} becomes: 

\begin{equation}
\frac{d C_2}{dt} + \frac{V_s}{h_2} C_2 = 0.
\label{mass_bal12}
\tag{S35}
\end{equation}

Assuming that all particles have crossed the interface by the time $t_{lim}$, we substitute $t_{lim}$ into \ref{mass_bal11} in order to find the initial condition for equation \ref{mass_bal12}. Thus the initial condition is now:

\begin{equation}
C_2 \left( t_{lim} \right) = C_1 \left( 0 \right) \left[ 1 + \left( \frac{V_s}{h_2} t_a - 1 \right) e^{- \frac{V_s}{h_2} \left( t_{lim} - t_a \right)} \right].
\label{mass_bal13}
\tag{S36}
\end{equation}
which allows to solve equation \ref{mass_bal12} and to find the solution of $C_2 \left( t \right)$ for a convective lower layer at $t>t_{lim}$

\begin{equation}
C_2 \left( t \right) = C_1 \left( 0 \right) e^{\frac{h_1}{h_2}} \left[ 1 + \left( \frac{V_s}{h_2} t_a - 1 \right) e^{- \frac{V_s}{h_2} \left( t_{lim} - t_a \right)} \right] e^{- \frac{V_s}{h_2} t}.
\label{mass_bal14}
\tag{S37}
\end{equation}

On the other hand, if we assume that the convection stops at $t_{lim}$, then we have to consider a quiescent settling in the lower layer. Thus the associated constant flux becomes

\begin{equation}
\dot{M}_{out} = A V_s C_2 \left( t_{lim} \right),
\label{mass_bal15}
\tag{S38}
\end{equation}
and \ref{mass_bal12} becomes

\begin{equation}
\frac{d C_2}{dt} + \frac{V_s}{h_2} C_2 \left( t_{lim} \right) = 0.
\label{mass_bal15}
\tag{S39}
\end{equation}

Then, keeping the initial condition \ref{mass_bal13} to solve \ref{mass_bal12}, the solution of $C_2 \left( t \right)$ for a quiescent lower layer at $t>t_{lim}$ is

\begin{equation}
C_2 \left( t \right) = C_1 \left( 0 \right) \left[ 1 + \left( \frac{V_s}{h_2} t_a - 1 \right) e^{- \frac{V_s}{h_2} \left( t_{lim} - t_a \right)} \right] \left( 1 + \frac{h_1}{h_2} - \frac{V_s}{h_2} t \right).
\label{mass_bal16}
\tag{S40}
\end{equation}

\end{itemize}

\subsection{Mass of particles accumulating at the bottom}

From the mass balance models, \cite{Hoyal1999} derived an estimate for the mass of particle which accumulate at the bottom of the domain. Here we detail the calculation of the mass and extend it by including the solution of $C_2 \left( t \right)$ for $t>t_{lim}$ (presented above).
Particles start to accumulate as soon as they reach the bottom of the tank (i.e. after a time $t_a$). Then for $t<t_a$, we have:

\begin{equation}
m_b = 0.
\label{mass1}
\tag{S41}
\end{equation}
where $m_b$ is the mass at the base of the tank. Furthermore, for $t_a \leq t < t_{lim}$ we integrate the flux $A V_s C_2 \left( t \right)$ between $t_a$ and $t$, using equation \ref{mass_bal11} for $C_2 \left( t \right)$. Thus we have:

\begin{equation}
m_b = m_0 \frac{V_s}{h_1} \left[ t + \left( \frac{h_2}{V_s} - t_a \right) e^{- \frac{V_s}{h_2} \left( t - t_a \right)}  - \frac{h_2}{V_s} \right],
\label{mass2}
\tag{S42}
\end{equation}
where $m_0$ is the initial mass of particles introduced in the upper layer. We observe that this relation is equivalent to equation 24 derived in \cite{Hoyal1999}, but delayed by $t_a$. Finally, for $t>t_{lim}$ we integrate the flux between $t_{lim}$ and $t$, taking into account that some particles have already accumulated between $t_a$ and $t_{lim}$ according to \ref{mass2}. As we have presented two different cases for the lower layer after $t_{lim}$ (convective and quiescent), there are also two possibilities for the mass of particles. Then, for a convective lower layer we have:

\begin{equation}
m_b = m_0 \frac{V_s}{h_1} \left\{ t_{lim} - \frac{h_2}{V_s} e^{\frac{h_1}{h_2}} \left[ 1 + \left( \frac{V_s}{h_2} t_a - 1 \right) e^{- \frac{V_s}{h_2} \left( t_{lim} - t_a \right)} \right] e^{- \frac{V_s}{h_2} t} \right\},
\label{mass3}
\tag{S43}
\end{equation}
and for a quiescent lower layer:

\begin{equation}
m_b = m_0 \frac{V_s}{h_1} \left\{ t_{lim} - \frac{h_2}{V_s} \left[ 1 + \left( \frac{V_s}{h_2} t_a - 1 \right) e^{- \frac{V_s}{h_2} \left( t_{lim} - t_a \right)} \right] \left( 1 + \frac{h_1}{h_2} - \frac{V_s}{h_2} t \right) \right\} ,
\label{mass4}
\tag{S44}
\end{equation}

%\begin{table}[]
%\centering
%\begin{tabular}{|c|l|}
\begin{longtable}{| p{.20\textwidth} | p{.80\textwidth} |}
\hline
$A$ & Area of the plane through which the particle mass flux is computed  \\
$b$ & Coefficient used to compute the force term in the \cite{Guo2002} formulation  \\
$c_i$ & Local particle velocity associated to a given lattice type in the LBM  \\
$c_s$ & Speed of sound used in the LBM \\
$C_1$ & Particle concentration in the upper layer \\
$C_2$ & Particle concentration in the lower layer \\
$D, D_c, D_S$ & Diffusion coefficients respectively for the density-altering quantity, the particle field and the sugar \\
$D_p$ & Particle diameter\\
$D_z$ & Differential operator \\
$\vec{e_z}$ & Vertical unit vector \\
$f_i, f_i^{eq}$ & Particle population and equilibrium distribution function\\
$\vec{F}$ & Body force term \\
$F_i$ & Power series expansion in the \cite{Guo2002} formulation \\
$Fr$ & Froude number \\
$g, g'$ & Gravitational acceleration and reduced gravity \\
$Gr_c$ & Critical Grashof number\\
$h_1,h_2$ & Thicknesses of the upper and lower layer respectively\\
$H,H_o$ & Heights of the particle interface at any time and at time $t=0$ \\
$\tilde{H}, |\tilde{H_i}|$ & Fourier transform of $H$ and the associated initial amplitude \\
$I$ & Identity operator \\
$J$ & Particle mass flux \\
$k,k_{sim}$ & Wavenumber associated with the instability and the value computed from the simulations \\
$k_S$ & Sampling wavenumber \\
$K$ & Left term in the matrix form of the eigenvalue problem in the LSA \\
$l^c$ & Characteristic length used in the viscous scaling \\
$l_x,l_y,l_z$ & Domain extent respectively in the $x$,$y$ and $z$ directions \\
$L$ & Characteristic length of the flow \\
$L_S$ & Number of samples in the instability Fourier analysis \\
$m_b$ & Mass of particles accumulated at the bottom of the domain \\
$M$ & Compacted form of the second order differential operator $(M = -k^2 + D_z^2)$ \\
$M_2$ & Mass of particles in the lower layer \\
$\dot{M}_{in},\dot{M}_{out}$ & Particle mass flux entering and leaving the lower layer respectively \\
$NE$ & Numerical diffusion error in the first order finite-difference scheme \\
$p,p^c$ & Fluid pressure and characteristic pressure in the viscous scaling \\
$q$ & Power used in the law defining $V_f$ as a function of $\phi$ \\
$S,S_0,S^{*}, \bar{S},\hat{S}$ & Sugar concentration, sugar concentration at time $t=0$ and $S^{*} (=S_0)$ is used in the nondimensionalisation, base state and perturbation amplitude for the sugar concentration \\
$Sc_i$ & Schmidt number \\
$St$ & Stokes number \\
$t,t_a,t^c,t_{lim}$ & Time, time for the particle first arrival at the bottom of the tank, characteristic time in the viscous scaling and time at which all particle have settle across the initial interface. \\
$T$ & Time at which the instability starts growing \\
$u,\vec{u_f}$ & Transport velocity and fluid velocity \\
$U$ & Characteristic velocity \\
$ V_f,V_s$ & Respectively the fingers velocity and individual particle settling velocity \\
$W$ & Right term in the matrix form of the eigenvalue problem in the LSA \\
$\vec{x}=(x,y,z)$ & Position vector and associated 3D components \\
$z_{\phi},z_s$ & Error function parameters used for the base states of volmue fraction and sugar respectively \\
$\alpha$ & Sugar expansion coefficient \\
$\Gamma_H, \Gamma_{H_i}$ & Power spectral density (PSD). The $i$ index stands for the initial PSD \\
$\delta t,\delta x$ & Temporal and spatial steps \\
$\delta_{PBL}$ & PBL thickness \\
$\Delta t$ & Integration time for the particle mass flux \\
$\eta$ & Power used in the law defining $V_f$ as a function of $D_p$ \\
$\mu$ & Fluid dynamic viscosity \\
$\nu$ & Fluid kinematic viscosity \\
$\rho,\rho_0,\rho_f,\rho_p,\rho_{PBL},\rho_{blk}$ & Density, fresh water density, fluid density (including sugar), particle density, PBL density, bulk density (including sugar and particles) \\
$\sigma,\sigma_{sim},\sigma_{LSA,i}$ & Growth rate of the instability, growth rate computed from simulations and growth rate predicted by LSA \\
$\tau$ & Relaxation coefficient in the LBM \\
$\varphi,\bar{\varphi},\varphi',\hat{\varphi}$ & Arbitrary variable, associated base state, perturbation and perturbation amplitude \\
$\phi, \phi_{tot},\phi_i,\phi_0,\phi^*,\bar{\phi},\hat{\phi}$ & Particle volume fraction, total particle volume fraction (polydisperse case), volume fraction of the $i-th$ size class, initial particle volume fraction, $\phi^{*}$ ($=\phi_0$) is used in the nondimensionalisation, volume fraction base state, perturbation amplitude \\
$\psi, \bar{\psi}, \hat{\psi}$ & Stream function, associated base state, perturbation amplitude \\
$\omega, \bar{\omega},\hat{\omega}$ & Vorticity, associated base state, perturbation amplitude s\\
\hline
%\end{tabular}
%\end{table}
\caption{List of symbols used in the main manuscript}
\label{tab1}
\end{longtable}

\bibliographystyle{apalike}
\bibliography{library}

%----------------------------------------------------------------------------------------